\documentclass[showpacs,aps,prd,nofootinbib,floatfix,amsmath,amssymb]{revtex4}
\usepackage{graphicx}
\usepackage{enumerate}
\usepackage{amsmath}
\usepackage{amssymb}
\usepackage{amsfonts}
\usepackage{color}
\usepackage{pstricks}
\usepackage{color}
\usepackage{subfig}
\usepackage{multirow}
\usepackage[utf8x]{inputenc}
\begin{document}
\title{\boldmath SM Higgs boson and $t\rightarrow cZ$ decays in the 2HDM type III with CP violation}
\author{R. Martinez}
\email{remartinezm@unal.edu.co} 
\affiliation{Departamento de F\'isica, Universidad Nacional de Colombia, Bogot\'a D.C., Colombia}
\author{R. Gait\'an}
\email{rgaitan@unam.mx} 
\affiliation{Departamento de F\'isica, FES-Cuautitl\'an, UNAM, C.P. 54770, Estado de M\'exico, M\'exico}
\author{J.H. Montes de Oca}
\email{josehalim@comunidad.unam.mx}
\affiliation{Departamento de F\'isica, FES-Cuautitl\'an, UNAM, C.P. 54770, Estado de M\'exico, M\'exico}
\author{E. A. Garc\'es}
\email{estela.garces@gmail.com} 
\affiliation{Departamento de F\'isica, Centro de Investigaci\'on y Estudios Avanzados del IPN, Ciudad de M\'exico, M\'exico}

\begin{abstract}
We compute the contributions  to rare top decays $t\rightarrow cZ$ and $t\rightarrow ch_1$ from the scalar sector in the  2HDM type III with CP violation, where $h_1$ is the Standard Model Higgs boson.
The branching ratio for $BR(t\rightarrow c Z)$ and $BR(t\rightarrow c  h_1)$ are obtained as a function of the model parameters. In particular, the $BR(t\rightarrow c Z)$ can increase its value up to $10^{-3}$ for $\tan\beta= 1$ and masses for the additional Higgs bosons of $m_{h_2,h_3,H^\pm}\sim 0.5$ TeV. 
Meanwhile $BR(t\rightarrow c h_1)$ can reach values of the order of $\sim10^{-2}$.  
We constrain the model parameters (mixing angles of the neutral scalar fields in the CP violation context  and $\tan\beta$) using the reported values of the signal strengths $R_{XX}$ and $b \rightarrow s \gamma$ process. 

\end{abstract}
%
\maketitle
%
%
\section{Introduction}
\label{sec1}
One of the goals of the Large Hadron Collider (LHC)  was  observing the Higgs Boson and to looking for  physics beyond the Standard Model (SM). 
In 2012 the ATLAS and CMS collaborations took a big step with the observation of a SM-like Higgs boson with a mass of 125 GeV~\cite{Chatrchyan:2012xdj,Aad:2012tfa}; 
nevertheless it was the first step in the long search of the Higgs boson from its theoretical assumption by the Standard Model (SM). 
This theory originally incorporated only one Electroweak (EW) doublet scalar field where the Higgs boson particle arises when the symmetry $SU(2) \otimes U(1)$ is broken. 
Currently, there are no experimental and theoretical restrictions to suppose only one EW doublet scalar field, 
which suggests to consider models with more scalar fields in order to study physics Beyond Standard Model (BSM). 
One of the simplest models reported in the literature is the Two Higgs Doublet Model (2HDM)~\cite{Gunion:1989we,Glashow:1976nt,Atwood:1995ej,Atwood:1995ud,Atwood:1996vw,Abbas:2015cua,Atwood:1996vj,Sher:1991km,Cheng:1987rs} which explains the hierarchy between the quark masses in the different families as a consequence of the hierarchy of Vacuum Expectation Values (VEVs). 
 The classification of the 2HDM is reviewed in detail in the report~\cite{Branco:2011iw}. 
 In the literature, usually the discrete symmetry $Z_2$ 
 is used to control the couplings and the models are classified according to their assignment of the values for the $Z_2$ charges in doublets  and fermions. For instance, for the model named as 2HDM type I only one of the doublets give masses to the fermions~\cite{Haber:1978jt}, 
 while in the 2HDM type II both doublets participate in the masses of the fermions where each doublet is assigned to give mass to each fermion sector, respectively. One for the up and the other for the down sector\cite{Donoghue:1978cj}. Without this $Z_2$ symmetry both doublet scalar fields give masses to the up and down sectors (Type-III).

There are many motivations to extend the scalar sector of the SM. To understand the relic density of Dark Matter (DM) of the Universe, one possibility, 
is the introduction of one scalar singlet which must have a vacuum expectation value (VEV) equal to zero, this in order to avoid faster decay in SM particles and have the abundance according to the stelar dynamic and lensing effects\cite{McDonald:1993ex,McDonald:2001vt,LopezHonorez:2006gr}.

 Another possibility  for the introduction of DM candidates is to add an EW doublet scalar field with VEV equal to zero; this model is known as  the Inert Doublet Model (IDM). Another interesting motivation for the extension of the scalar sector is the inclusion of CP- violation in order to incorporate leptogenesis and the matter antimatter excess in the Universe~\cite{Sakharov:1967dj,Cosme:2005sb,Fuyuto:2017ewj}. 

Neutrino oscillations~\cite{Pontecorvo:1957cp,Pontecorvo:1957qd,Maki:1962mu}, transitions between different neutrino flavors $\nu_e$, $\nu_{\mu}$, $\nu_{\tau}$, caused by nonzero neutrino masses, have been observed  in the experiments with solar, atmospheric, reactor and accelerator neutrinos ~\cite{Cleveland:1998nv,Fukuda:1996sz,Anselmann:1992um,Ahmad:2002jz,Fukuda:2002pe,Eguchi:2002dm,Araki:2004mb,Fukuda:1998mi,Ashie:2005ik}. The observation of neutrino oscillation requires  the neutrino masses to be incorporated in the SM~\cite{Froggatt:1997he}. In order to introduce masses to neutrinos an interesting proposal is to include right-handed or sterile neutrinos. New scalar fields are also necessary to generate masses through see-saw mechanism, which can be an EW doublet in order to generate Dirac masses or an EW singlet to give Majorana masses. As a consequence, we have to introduce a unitarity matrix which relates the mass and flavor eigenstates. 
The diagonalization of the mass matrices of charged leptons and neutrinos generates the so called Pontecorvo-Maki-Nakagawa-Sakata (PMNS)  mixing matrix~\cite{Pontecorvo:1957cp,Pontecorvo:1957qd,Maki:1962mu} . The PMNS matrix works similarly as the CKM matrix, plus two additional CP phases for Majorana neutrinos. CP-violation has been measured in the quark sector for the system $K^0-\bar{K}^0$ and $B^0_{s(d)}-\bar{B}^0_{s(d)}$ ~\cite{Grossman:1996era,Dunietz:2000cr,Langacker:2000ju,Barger:2003hg,Fajfer:2001ht,Perez:1992hc,Anderson:2005ab,Rodriguez:2004mw,Promberger:2007py,Barger:2004qc,Cheung:2006tm,Grossman:2006ce,Martinez:2008jj}. On the other hand, long baseline neutrino experiments, like NOvA, T2K and Minos have observed CP-violation in the neutrino sector~\cite{Adamson:2016tbq},~\cite{GonzalezGarcia:2007ib}. 

The study of models with new sources of CP-violation is very well motivated. In particular, the 2HDM type III, without including the $Z_2$ discrete symmetry, allows CP-violation simultaneously in the scalar sector~\cite{Basso:2012st} and in the Yukawa Lagrangian. Under this assumption, the neutral CP-even and CP-odd Higgs bosons are combined in a scalar-pseudoscalar structure and their mass eigenstates do not have defined CP-parity. The pseudoscalar coupling depends of the CP-violation of the model and it must be strongly suppressed. 

Indirect evidence of a new physics signal in rare processes mediated by Flavor-Changing Neutral Currents (FCNC) could give a crucial direction to BSM physics~\cite{Lindner:2016bgg,Hall:1981bc}. The main motivation for considering FCNC is that these processes are extremely suppressed in the SM while their extensions are improved by FCNC approaching the experimental limits. The rare decays with FCNC, which have the greatest increase, are associated with the top quark such as $t\rightarrow q   V$ for $q=u,c$ and $V=\gamma, g, Z$~\cite{Eilam:1990zc,AguilarSaavedra:2002ns,AguilarSaavedra:2004wm,Mele:1998ag,DiazCruz:1989ub,Larios:2006pb}. The current experimental limit is yet 10 orders of magnitude apart from SM; SM value is of the order of $10^{-17}\sim10^{-12}$ ~\cite{Eilam:1990zc,AguilarSaavedra:2002ns,AguilarSaavedra:2004wm,Mele:1998ag,DiazCruz:1989ub,Larios:2006pb} meanwhile current limit is $BR<5.6\times10^{-3}$~\cite{Patrignani:2016xqp}.  In 2HDM with CP conserving the rare top decays present an increase in the branching ratio  the order of $10^{-7}-10^{-9}$\cite{Atwood:1996vj,Grzadkowski:1990sm,Arhrib:2005nx,Bejar:2000ub,Luke:1993cy,Atwood:1995ud,Atwood:1995ej,Abbas:2015cua}. This means that a signal of rare top decays near the LHC experimental limits will be a clear evidence of new physics~\cite{Gaitan:2015hga,Diaz-Furlong:2016ril,Hesari:2015oya,Enomoto:2015wbn,Dey:2016cve,Khatibi:2015aal,Gaitan:2017cfa,Bardhan:2016txk}. We analyze the FC in the context of 2HDM type III with CP violation which will introduce parameters, such as of $\alpha_1$, $\alpha_2, \alpha_3$, and they are absent in the usual models. In order to find allowed regions for $\left\{  \alpha_1,\alpha_2,\tan\beta \right\}$ and $Y_{tc}$, we consider the contributions of the pseudoscalar couplings between the fermions and $h_1$ to $R_{XX}$ by using the LHC measurements. Then we  will find the values for the $BR(t\to cZ)$ and $BR(t\to c h_1)$ for this region of parameters.

In section \ref{sec2} we present the model. In section \ref{sec3} we find the allowed region for the parameter space based on experimental values and $\chi^2$ analysis. The section \ref{sec4} is devoted to present our results for the rare top decay. Finally in section \ref{sec5} we discuss the obtained result and the perspectives for the model and the conclusions in section \ref{sec6}. 
%
%
\section{Mixing and Flavor-Changing neutral scalars in 2HDM}
\label{sec2}

Let us denote the two complex $SU(2)_L$ doublet scalar fields with hypercharge 1 as $\Phi_1$ and $\Phi_2$.  If the $\Phi_{1,2}$ are included in the most general form in the scalar potential and Yukawa interactions, FC through neutral scalar fields can arise with the fermion interactions and a general mixing for the three physical states of the neutral  scalar. Usually, the discrete symmetry $Z_2$ is introduced in order to suppress these features in the model. This suppression is motivated by the experimental limits for FC processes, however it could give signs of new physics and CP violation effects. 

In the 2HDM, one linear combination of the Yukawa couplings  is proportional to the mass fermion and this can be diagonalized by a bi-unitarity matrix. The other linear combination cannot be simultaneously diagonalized and this new coupling produces flavor change (FC).  This kind of model is the so called 2HDM type III. In the scalar potential appear new bilinear and quartic interactions like $\Phi_1^\dagger \Phi_2$ and $\Phi_1^\dagger \Phi_1 \Phi_1^\dagger\Phi_2$ which can induce CP violation explicitly. We study the 2HDM type III with explicit CP violation and FCNSI which is described below.

In the 2HDM type III the mixing of the neutral Higgs bosons, usually denoted as $h^0$, $H^0$ and $A^0$, can be parametrized by the three angles $\{ \alpha_1,\alpha_2, \alpha_3\}$~\cite{ElKaffas:2006gdt}, but in all decay channels of the lightest neutral Higgs boson, $h_1^0$, only $\alpha_1$ and $\alpha_2$ are required to describe the width decays. On the other hand, if $\alpha_2=0$, the mixing of the neutral Higgs bosons recover the CP-parity and  $\alpha_1$ is the usual angle that mixes $h^0-H^0$ in the CP-conserving 2HDM. In this scenario $\alpha_2$ is an important parameter to analyze CP-violation because when it is zero all analytical expressions must be reduced to 2HDM type I or II. Using the signal strengths, $R_{XX}$, reported by LHC~\cite{Patrignani:2016xqp}  and the predicted by 2HDM-III, we will find the allowed region for $\alpha_1$ and $\alpha_2$ as a function of $\tan\beta$,  defined as  a ratio of VEVs. 
%
%
\subsection{Yukawa interactions with FC}
\label{subsec:2-1}
The most general structure for the Yukawa couplings among fermions and scalars is
\begin{equation}
\mathcal{L}_{Y}=\sum_{i,j=1}^{3}\sum_{a=1}^{2}\left( \overline{q}
_{Li}^{0}Y_{aij}^{0u}\widetilde{\Phi }_{a}u_{Rj}^{0}+\overline{q}
_{Li}^{0}Y_{aij}^{0d}\Phi _{a}d_{Rj}^{0}+\overline{l}_{Li}^{0}Y_{aij}^{0l}
\Phi _{a}e_{Rj}^{0}+h.c.\right) ,  \label{yukawa}
\end{equation}
where $Y_{a}^{u,d,l}$ are the $3\times 3$ Yukawa matrices. $q_{L}$ and $l_{L}$
denote the left handed fermion doublets under $SU(2)_L$, while $u_{R}$, $d_{R}$, $l_{R}$ correspond to the right handed singlets. 
The zero superscript in fermion fields and Yukawa matrices stands for the interaction basis and non diagonal matrices in the most general case, 
respectively. The doublets are written as 
\begin{equation}
 \Phi_{a}=\left(
\begin{array}{c}
\phi_{a}^{+} \\
\phi_{a}^0\\
\end{array}
\right)
a = 1, 2
\label{phis}
\end{equation}
The relation between the interaction and physical states is found through the Spontaneous Symmetry Breaking (SSB), 
where the most general $U(1)_{EM}$-conserving VEVs can be taken as
\begin{equation}
\langle \Phi_1 \rangle= \frac{1}{\sqrt{2}}\left(
\begin{array}{c}
0 \\
v_1 \\
\end{array}
\right),
\label{vev1}
\end{equation}
\begin{equation}
\langle \Phi_2 \rangle= \frac{1}{\sqrt{2}}\left(
\begin{array}{c}
0 \\
v_2\\
\end{array}
\right),
\label{vev2}
\end{equation}
$v_1$ and $v_2$ are real and satisfy $v^2 \equiv v_1^2 + v_2^2 = \frac{4 M_W^2}{g^2} $\cite{Ginzburg:2004vp}.
After getting a correct SSB,  the Eq.(\ref{vev1}) and Eq.(\ref{vev2}) are used in Eq. (\ref{yukawa}) to obtain the mass matrices which are written as
\begin{equation}
M^{u,d,l}=\sum_{a=1}^{2}\frac{v_{a}}{\sqrt{2}}Y_{a}^{u,d,l},  \label{mass}
\end{equation}
where $M^{u,d,l}=\textrm{diagonal}\left\{m_{u,d,e},\,m_{c,s,\mu},\,m_{t,b,\tau}\right\}$ and $Y_a^{f}=V_L^f Y_a^{0f}\left(V_R^{f}\right)^\dag$, for $f=u,d,l$. The $V_{L,R}^f$ matrices are used to diagonalize the fermion mass matrices and to relate the physical and interaction states for fermions. Note that in 2HDM-III the diagonalization of mass matrices does not imply the diagonalization of the Yukawa matrices, as it occurs in the 2HDM type I or II. An important consequence of non diagonal Yukawa matrices in physical states is the presence of FCNSI between neutral Higgs bosons and fermions.

We will only focus in the quarks, however the charged leptons can be included in an analogous form. The equations (\ref{mass}) not only establish the mass matrices but also provide relations to eliminate one of the Yukawa matrices in the physical states. In order to obtain the interactions in terms of only one Yukawa matrix, the equations (\ref{mass}) can be written in two possible forms 
\begin{equation}
Y_1^{q}=\frac{\sqrt{2}}{v_1}M^{q}-\frac{v_2}{v_1}Y_2^{q}
\label{y1}
\end{equation}
or
\begin{equation}
Y_2^{q}=\frac{\sqrt{2}}{v_2}M^{q}-\frac{v_1}{v_2}Y_1^{q},
\label{y2}
\end{equation}
where the quark sector label is $q=u,d$. The VEV's ratio defines the $\beta$ parameter, $\tan\beta=\frac{v_2}{v_1}$, then $v_1=v\sin\beta$ and $v_2=v\cos\beta$. By using
Eq. (\ref{y1}) or Eq. (\ref{y2}) in the Yukawa Lagrangian $\mathcal{L}_Y$ , Eq. (\ref{yukawa}), the 2HDM type III can be written in four different versions.

From Eq. (\ref{y1}) and Eq. (9) we can find $Y_2^U$ and $Y_2^D$ as a function of the other Yukawas and masses obtaining:
\begin{equation}
Y_2^{U}=\frac{\sqrt{2}}{v_2}M^{u}-\frac{v_1}{v_2}Y_1^{u}, \;\;
Y_2^{D}=\frac{\sqrt{2}}{v_2}M^{d}-\frac{v_1}{v_2}Y_1^{d}.
\end{equation}
Replacing them into eq. (\ref{yukawa}), we obtain the Lagrangian 2HDM type I plus FC interactions. On the other hand, from eqs.(6-7)
we can also solve for
\begin{equation}
Y_2^{U}=\frac{\sqrt{2}}{v_2}M^{u}-\frac{v_1}{v_2}Y_1^{u},\;\;
Y_1^{D}=\frac{\sqrt{2}}{v_1}M^{d}-\frac{v_2}{v_1}Y_2^{d}.
\end{equation}
Replacing them into eq. (\ref{yukawa}), we obtain the Lagrangian 2HDM type II plus FC interactions.
There are other two different 2HDM Lagrangians and they are combinations of the type I and II. They can be obtained by solving the Yukawas 
in the following form:
\begin{equation}
Y_1^{u}=\frac{\sqrt{2}}{v_1}M^{U}-\frac{v_2}{v_1}Y_2^{u},\;\;
Y_1^{d}=\frac{\sqrt{2}}{v_1}M^{D}-\frac{v_2}{v_1}Y_2^{d}.
\end{equation}
and 
\begin{equation}
Y_1^{u}=\frac{\sqrt{2}}{v_1}M^{U}-\frac{v_2}{v_1}Y_2^{u},\;\;
Y_2^{d}=\frac{\sqrt{2}}{v_2}M^{D}-\frac{v_1}{v_2}Y_1^{d}.
\end{equation}
Taking into account similar rotations for the lepton sector there are only two Feynman rules which correspond to 2HDM type I and II plus FC. 
The general structure for the interactions between the quarks and neutral scalars in any of them is
\begin{equation}
\bar{q}\left[f(\beta)(A_LP_L+A_RP_R)M^q-g(\beta)(B_LP_L+B_RP_R)Y^q\right]q h_k,
\end{equation}
where $P_{R,L}=\frac{1}{2}\left(1\pm \gamma_5\right)$.  The $f(\beta)$ or $g(\beta)$ can be written as sine, cosine, tangent or cotangent of $\beta$, which will depend on the model version. The $A_{L,R}$ contain all the information related with the mixing of the neutral scalars $h_k$,  which will be discussed below. The mass matrix must be diagonal, meanwhile the Yukawa matrix could be, in general, non diagonal. These elements of the Yukawa matrix are responsible for the FC mediated by neutral scalars.
\subsection{Neutral scalar mixing from the scalar potential}
\label{subsec:2-2}
Given $\Phi_1$ and $\Phi_2$ two complex $SU(2)_L$ doublet scalar fields, the most general gauge invariant and renormalizable Higgs scalar potential is
\cite{Haber:1993an,Xu:2017vpq}
\begin{eqnarray}
V &=&m_{11}^{2}\Phi _{1}^{+}\Phi _{1}+m_{22}^{2}\Phi _{2}^{+}\Phi _{2}-\left[
m_{12}^{2}\Phi _{1}^{+}\Phi _{2}+h.c.\right] +\frac{1}{2}\lambda _{1}\left(
\Phi _{1}^{+}\Phi _{1}\right) ^{2}  
+ \frac{1}{2}\lambda _{2}\left( \Phi _{2}^{+}\Phi _{2}\right) ^{2} \nonumber \\ 
&& + \lambda_{3}\left( \Phi _{1}^{+}\Phi _{1}\right) \left( \Phi _{2}^{+}\Phi_{2}\right) 
  +\lambda _{4}\left( \Phi _{1}^{+}\Phi _{2}\right)  \left( \Phi_{2}^{+}\Phi _{1}\right)  
  + \frac{1}{2}\lambda _{5}\left (\left( \Phi _{1}^{+}\Phi _{2}\right)^{2} + \left( \Phi _{2}^{+}\Phi _{1}\right)^{2}\right)
  \nonumber \\ 
  &&
  +\left[\lambda _{6}\left( \Phi _{1}^{+}\Phi _{1}\right) \left( \Phi
_{1}^{+}\Phi _{2}\right) +\lambda _{7}\left( \Phi _{2}^{+}\Phi _{2}\right)
\left( \Phi _{1}^{+}\Phi _{2}\right)  + h.c. \right] ,
\end{eqnarray}
where $m_{11}^2$, $m_{22}^2$ and $\lambda_1$, $\lambda_2$, $\lambda_3$, $\lambda_4$, $\lambda_5$ are real parameters and $m_{12}^2$, $\lambda_6$, $\lambda_7$ can be complex parameters. 
The neutral components of the doublets in the interaction basis are $\phi_a^0=\frac{1}{\sqrt{2}}\left( v_a + \eta_a +i \chi_a\right)$, where $a=1,2$. As a result of the explicit CP symmetry breaking introduced, Eq. (13), a mixing matrix $R$ relates the mass eigenstates $h_i$ with the $\eta_i$ as follows
\begin{equation}
h_{i}=\sum_{j=1}^{3}R_{ij}\eta _{j}.
\label{h-Rn}
\end{equation}
$R$ matrix is parametrized in the usual form as~\cite{ElKaffas:2006gdt}:
\begin{equation}
R=\left(
\begin{array}{ccc}
c_{1}c_{2} & s_{1}c_{2} & s_{2} \\
-\left( c_{1}s_{2}s_{3}+s_{1}c_{3}\right)
& c_{1}c_{3}-s_{1}s_{2}s_{3} & c_{2}s_{3} \\
-c_{1}s_{2}c_{3}+s_{1}s_{3} & -\left(
c_{1}s_{3}+s_{1}s_{2}c_{3}\right)  &
c_{2}c _{3}
\end{array}
\right),
\label{r_matrix}
\end{equation}

Here, $\eta_3$ is the state orthogonal to the would-be Goldstone boson assigned to the $Z$ gauge boson, explicitly it is written as $\eta_3=-\chi_1 $ $\sin\beta+\chi_2\cos\beta$, 
where $c_i=\cos\alpha_i$, $s_i=\sin\alpha_i$ for $-\frac{\pi}{2}\leq\alpha_{1,2}\leq\frac{\pi}{2}$ and $0\leq\alpha_3\leq\frac{\pi}{2}$. $h_i$ satisfy the mass relation $m_{h_1}\leq m_{h_2}\leq m_{h_3}$~\cite{Basso:2012st,Arhrib:2010ju,Krawczyk:2013jta,Chen:2015gaa}. In the CP conserving case $\eta_{1}$ and $\eta_{2}$ are CP-even and mixed in a $2\times2$ matrix while $\eta_3$ is CP-odd decoupled from $\eta_{1}$ and $\eta_{2}$. However, due to CP-symmetry breaking, in general, the neutral Higgs bosons $h_{1,2,3}$ do not have well defined CP eigenstates.

The focus is on the up-type quark Yukawa interactions that contain the Feynman rules for the rare top decay. Replacing  Eq.(\ref{h-Rn}) and Eq.(\ref{mass}) in the Yukawa Lagrangian of Eq.(\ref{yukawa}), the interactions between neutral Higgs bosons and fermions can be written as interactions of the CP conserving 2HDM (type I or II) plus additional contributions, which arise from any of the $Y_{1,2}$ Yukawa matrices. The relation among the mass matrix $M^f$ and  the Yukawa matrices $Y_{1,2}^f$, for $f=u,d,l$, is used to write the Yukawa Lagrangian, Eq.(\ref{yukawa}), as a function  of only one Yukawa matrix, $Y_1^f$ or $Y_2^f$. We choose to write the interactions as a function of the Yukawa matrix $Y_2$, as follows,
\begin{equation}
Y_1^f= \frac{\sqrt{2}}{v_1}M^f-\frac{v_2}{v_1}Y_2^f.
\end{equation}
We will replace Eq. (16) in Eq.(\ref{yukawa}) for $f=u,d$. From now on, in order to simplify the notation, the subscript 2 in the Yukawa couplings will be omitted. Therefore, the interactions between quarks and neutral scalar bosons are explicitly written as
\begin{eqnarray}
\mathcal{L} &=&\frac{1}{v\cos \beta }\sum_{ik}\bar{u}_{i}M^{u}\left(
A_{k}P_{L}+A_{k}^{\ast }P_{R}\right) u_{i}h_{k}  
+\frac{1}{\cos \beta }\sum_{ijk}\bar{u}_{i}Y_{ij}^{u}\left(
B_{k}P_{L}+B_{k}^{\ast }P_{R}\right) u_{j}h_{k}   \nonumber \\
&&\frac{1}{v\cos \beta }\sum_{ik}\bar{d}_{i}M^{d}\left( A_{k}^{\ast
}P_{L}+A_{k}P_{R}\right) d_{i}h_{k} 
+\frac{1}{\cos \beta }\sum_{ijk}\bar{d}_{i}Y_{ij}^{d}\left( B_{k}^{\ast
}P_{L}+B_{k}P_{R}\right) d_{j}h_{k}  \nonumber \\
&& +\frac{1}{v\cos \beta }\sum_{ik}\bar{e}_{i}M^{l}\left( A_{k}^{\ast}P_{L}+A_{k}P_{R}\right) e_{i}h_{k}  +\frac{1}{\cos \beta }\sum_{ijk}\bar{e}_{i}Y_{ij}^{l}\left( B_{k}^{\ast
}P_{L}+B_{k}P_{R}\right) e_{j}h_{k} \nonumber \\
&&+\left[ \frac{\sqrt{2}}{\cos \beta }\sum_{ij}\bar{u}_{i}\left(
(KY^{d})_{ij}P_{R}-(Y^{u}K)_{ij}P_{L}\right) d_{j}H^{+}\right.   \nonumber \\
&&\left.+\frac{\sqrt{2}}{v}\tan \beta \sum_{ij}\bar{u}_{i}\left( -\left(KM^{d}\right) _{ij}P_{R}+\left( M^{u}K\right) _{ij}P_{L}\right) d_{j}H^{+}+h.c.\right],
\label{yukawa_quarks}
\end{eqnarray}
where we define
\begin{eqnarray}
A_k&=& R_{k1} - i R_{k3}\sin\beta ,\nonumber \\
B_k&=& R_{k2} \cos\beta - R_{k1} \sin\beta + i R_{k3}.
\label{ak}
\end{eqnarray}
The fermion spinors are denoted as $(u_1,\,u_2,\,u_3)=(u,\,c,\,t )$, where the indexes $i,\,j=1,2,3$ denote the family generations in Eq. (\ref{yukawa_quarks}), while $k=1,2,3$ is used for the neutral Higgs bosons. Note that a CP conserving case is obtained only if two neutral Higgs bosons are mixed with well-defined CP states, for instance $\alpha_2=\alpha_3=0$ is the usual limit.
%
%
\section{Constraints for FC neutral scalars}
\label{sec3}
Current observations in LHC impose restrictions on the neutral scalar $h_1$, which we will chose to be the SM Higgs boson. 
The strongest constraints for the $\alpha_{1,2}$ mixing angles come from its decay channels reported in the signal strength $R_{XX}$ for fermions and gauge bosons in the final state~\cite{Patrignani:2016xqp}. The first part of this section (A) is devoted to find bounds for $\tan\beta$ and charged Higgs mass using the measured $b\rightarrow s \gamma$  branching ratio. In the second part of this section, subsection B, we perform a $\chi^2$  analysis on $R_{XX}$ with statistical errors only.
We employ simultaneously  $R_{XX}$ and $b\rightarrow s \gamma$, and we to obtain an allowed region for the model parameters. 
%
%
\subsection{B physics constraints}
The FC decay of the bottom quark $b\rightarrow s \gamma$  imposes the strongest constraint on $\tan\beta$. In the 2HDM this decay has a one loop contribution from charged and neutral Higgs bosons. We will use the reported value of  $b\rightarrow s \gamma$ to constrain  $\tan\beta$ and Yukawa matrix element $Y_{tc}$. In order to find allowed values for the parameters we first review the possible constraints that $b\rightarrow s \gamma$ decay can impose on the $Y_{tc}$ coupling, assuming the charged Higgs in the range $500\sim 900$ GeV. New physics contributions can be parametrized in Wilson coefficients (WC). Following references~\cite{Degrassi:2000qf,Misiak:2006zs,Lunghi:2006hc,Gomez:2006uv,Barenboim:2013bla}, the branching ratio of the $b\rightarrow s \gamma$ decay is a function of the WC. The main contributions due to Wilson coefficients, beyond the charged current contribution, are given by the charged Higgs and FC Yukawa couplings, $\delta C_{7,8}=C^{H^{\pm}_{7,8}} +C^{H,FC}_{7,8}$. The charged-Higgs contribution is

\begin{equation}
C^{H^{\pm}_{7,8}} =\tan^2\beta\left( f_{7,8}^{(1)}(y_t) + f_{7,8}^{(2)}(y_t)\right),
\label{eq_h}
\end{equation}
while the FC contribution is 
\begin{eqnarray}
C_{7,8}^{H,FC} &=&\frac{2M_{W}}{gm_{t}K_{ts}\cos \beta }
(Y^{u}K)_{ts}f_{7,8}^{(1)}(y_{t}) 
+ \frac{2M_{W}}{gm_{b}K_{tb}\cos \beta }(KY^{d})_{tb}f_{7,8}^{(2)}(y_{t}),
\label{eq_chfc}
\end{eqnarray}
with $y_t=m^2_t/M^2_{H^\pm}$, the explicit relations for $f_{7,8}^{(1),(2)}(x)$ can be found in Ref.~\cite{Degrassi:2000qf,Misiak:2006zs,Lunghi:2006hc,Gomez:2006uv,Barenboim:2013bla}.
Using the hierarchy of the CKM matrix we have the following approximations $(Y^uK)_{ts}\approx Y_{tc}K_{cs}$ and $(KY^d)_{tb}\approx K_{tb}Y_{bb}$. In order to bound  the FC Yukawa coefficient $Y_{tc}$, we consider the Cheng-Sher Ansatz for $Y_{bb}$~\cite{Cheng:1987rs},  meaning $Y_{ij}\approx \sqrt{ m_i m_j}/v$, in particular $Y_{bb}\approx m_b/v$. 
Limits on the $B\to X_s\gamma$ decay come from the B factory experiments BaBar, Belle and CLEO~\cite{Chen:2001fja,Abe:2001hk,Lees:2012wg,Lees:2012ufa,Aubert:2007my}. The current HFAG world average for $E> 1.6$~GeV ~\cite{Amhis:2014hma}, is
\begin{eqnarray}
\textrm{BR} (B\to X_s\gamma) = (3.43 \pm 0.21 \pm 0.07) \times 10^{-4} .
\label{br_bsg}
\end{eqnarray}

\begin{figure}[htbp]
\centering 
\includegraphics[scale=0.7]{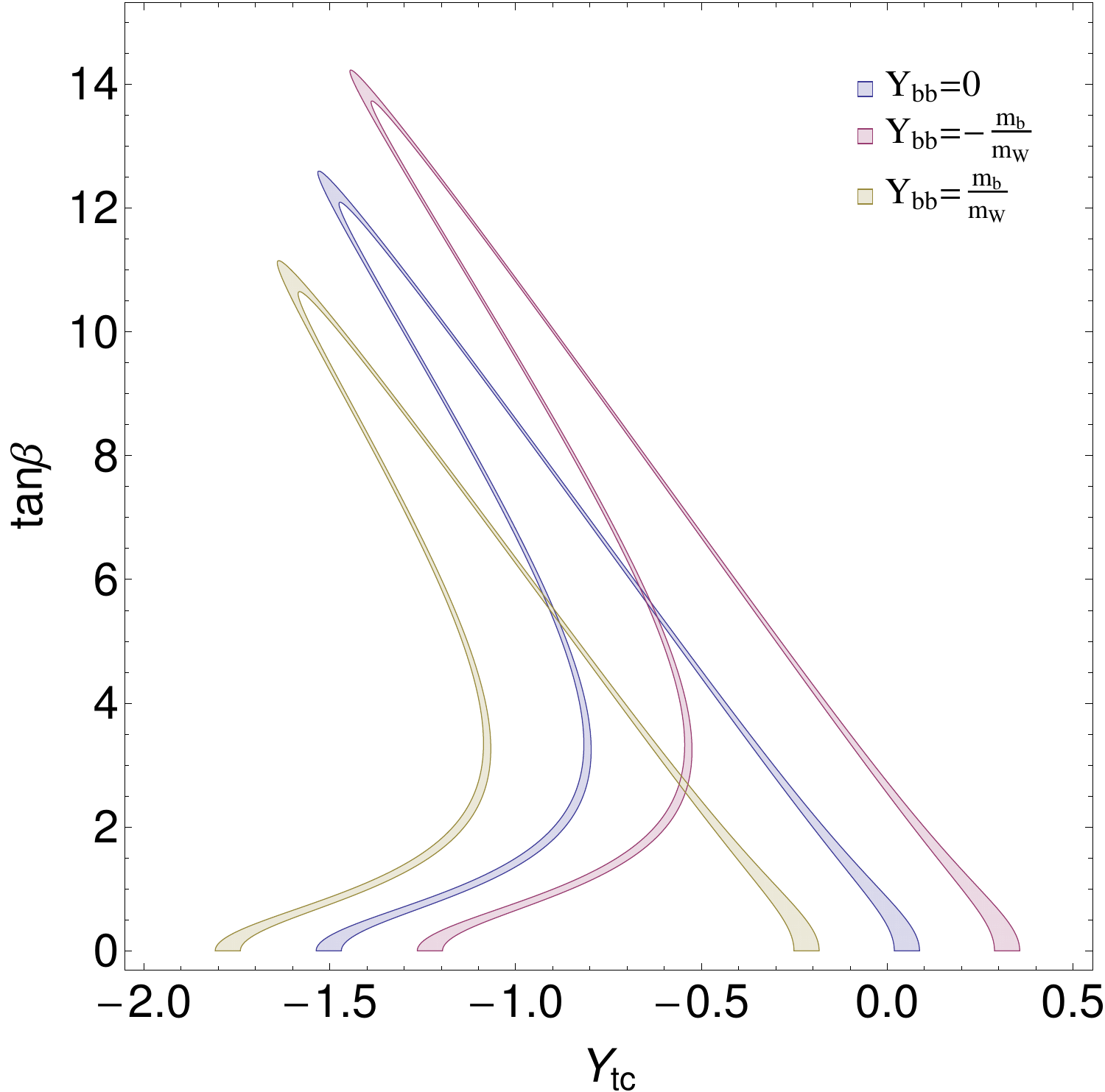}
\caption{\label{figurebsg} $\tan\beta$ as function of $Y_{tc}$ considering different values of $Y_{ff}$ and $m_{H^\pm}\approx 500$ GeV.}
\end{figure}

In figure \ref{figurebsg}, the Cheng-Sher Ansatz for three different values of the Yukawa couplings $Y_{bb}$ is considered to explore the allowed region in the  $\tan\beta$-$Y_{tc}$ plane. For  $Y_{bb}=0$ the allowed values for $Y_{tc}$ is around the zero value which reproduced the results for 2HDM with CP-conserving. 
%
%
%
%
\subsection{Constraints on the neutral scalar Higgs} 
The lightest neutral scalar $h_1$, Eq. (\ref{h-Rn}), is assumed to be the SM scalar $H^0$ observed by ATLAS and CMS collaborations~\cite{Aad:2012tfa, Chatrchyan:2012xdj}. The measured signal strength $R_{X\,X}$ for $X=b,W,Z,\tau,\gamma$ ~\cite{Patrignani:2016xqp}, see the table \ref{rxx}, can be used to constrain the parameter space in 2HDM. The signal strength for a given final state $XX$ is 
\begin{equation}
R_{XX}=\frac{\sigma\cdot BR(H^0\to X\,X)}{\left[\sigma\cdot BR(H^0\to XX)\right]_{SM}},
\end{equation}
where $H^0$ is the observed Higgs in pp collisions at LHC, channel $pp\to gg\to h\to\gamma\gamma$. Similarly, we can define a signal strength for the new neutral boson $h_1$ in the 2HDM as
\begin{equation}
R_{X\,X}^{2HDM}=\frac{\left[\sigma\cdot BR(h_1\to X\,X)\right]_{2HDM}}{\left[\sigma\cdot BR(H^0\to X\,X)\right]_{SM}}.
\end{equation}

The 2HDM is constructed in such a way that the quark-gluon interactions remain as in the SM. 
Therefore, the $R_{X\,X}^{2HDM}$ can be written only as the ratio of branching ratios of the 2HDM with CPV and SM multiplied by the ratio between the decay width $h_1\rightarrow g g$ and $H^0\rightarrow g g$, 
\begin{equation}
R_{X\,X}^{2HDM}\approx \frac{\Gamma\left(h_1\rightarrow g g\right)BR(h_1\to X\,X)}{ \Gamma\left(H^0\rightarrow g g\right)BR(H^0\to X\,X)}.
\label{road}
\end{equation}

We will use the limits on $R_{XX}$ reported by  ATLAS and CMS to constrain the parameters in the model, by using Eq. (\ref{road}). 

The $h_1$ with a mass of the order of $125$ GeV  decays at tree level in the channel $W^{+} W^{-}$, $Z Z$, $f \bar{f}$; for $f = e, \mu, \tau, b, c, s, d$, meanwhile the channels at one loop are $\gamma\gamma$, $\gamma Z$, $gg$.
A model with CPV as the 2HDM type III also introduces a pseudoscalar coupling between neutral Higgs and fermions which also appears inside the loop in  $h_1 \rightarrow \gamma\gamma, gg, \gamma Z$ through the top quark vertex. In general, since all fermions can give loop contributions since these contributions are proportional to the fermion masses, the top quark gives the greatest contribution.  

The decay widths in the 2HDM for $h_1\rightarrow f \bar{f}$, with $f=b,c,s,u,d,\tau,\mu,e$, are given by
\begin{equation}
\Gamma_{h_1\rightarrow f \bar{f}}=\frac{N_{c}m_{h_1}}{16\pi}\left(1-4\frac{m_f^2}{m_{h_1}^2}\right)^{\frac{1}{2}} \left[S^2_f\left(1-4\frac{m_f^2}{m_{h_1}^2}\right)+P^2_f \right],
\label{w_hff}
\end{equation}
where 
\begin{eqnarray}
\label{cf}
S_f&=&  \frac{m_f}{v\cos\beta}R_{11}+Y_{ff}\left( R_{12}-R_{11}\tan\beta \right), \nonumber\\
P_f&=& -\frac{m_f \tan\beta}{v}R_{13}+\frac{Y_{ff}}{\cos\beta}R_{13}.
\end{eqnarray}
$P_f$ is the additional contribution coming from pseudoscalar coupling between Higgs and fermions due to CPV.  For  $\alpha_2=0$, $P_f=0$. $S_f$ is the usual 2HDM contribution. For $f=b$ the reported result can be revised in~\cite{Kniehl:1994ju,Djouadi:1995gt,Chetyrkin:1995pd,Larin:1995sq,Melnikov:1995yp,Chetyrkin:1996sr,Spira:2016ztx,Kwiatkowski:1994cu}.

The Lagrangian used to calculate the Higgs decays at one loop level to $\gamma\,\gamma$, $\gamma\, Z$ and $gg$ is written as follows
\begin{equation}
\mathcal{L}=-\frac{g m_t}{2 m_W}\bar{u}(t)(a_f+i\tilde{a}_f\gamma_5)u(t)h_1+g m_W a_W W^{+}W^{-}h_1+\frac{1}{2}m_Z a_Z Z Z h_1-g\frac{m_{H^\pm}^2}{m_W}a_H H^+H^-h_1,
\label{lagrangianIII}
\end{equation}
where $a_{f,W,H}$ and $\tilde{a}_f$ are the deviations from the SM which are given by
\begin{eqnarray}
a_f&=&\frac{R_{11}}{\cos\beta}+\frac{Y_{tt}^u}{\cos\beta}\frac{2m_W}{gm_t}(R_{12}\cos\beta-R_{11}\sin\beta),\nonumber\\
\label{af}
\tilde{a}_f&=&(\sin\beta +\frac{Y_{tt}^u}{\cos\beta}\frac{2m_W}{gm_t})R_{13},\nonumber\\
a_W&=&a_Z=\cos\beta R_{11}+\sin\beta R_{12},\nonumber\\
a_H&=&1.
\label{coefficients-a}
\end{eqnarray}

It is important to note that if $\alpha_2=0$ then $\tilde{a}_f=0$. The three linear coupling in the charged and neutral Higgs bosons, $H^+H^-h_1$, can be approximated to $m_{H^\pm}^2$ in order to avoid the $\lambda_i$ parameters from the Higgs potential. In this case, we assume the scalar coefficient as $a_H=1$.

The partial width for the $h_1\rightarrow \gamma\gamma$  decay is given by ~\cite{Djouadi:2005gi,Spira:1995rr,Spira:1993bb,Djouadi:1990aj,Djouadi:1993ji,Liao:1996td,Martinez:1989bg,Martinez:1989kr,Martinez:1990ye}
\begin{equation}
\Gamma(h_1\rightarrow \gamma\gamma)=\frac{G_F\alpha^2m_{h_1}^3}{128\sqrt{2}\pi^3}\left[ \left| \frac{4}{3} a_t A_{1/2}(\tau_t)+a_W A_1(\tau_W)+A_0(\tau_{H\pm}) \right|^2 
+\left| \frac{8}{3} \tilde{a}_t \tilde{A}_{1/2}(\tau_t) \right|^2\right],
\label{h1gaga}
\end{equation}
where
\begin{eqnarray}
A_1(\tau)&=&2 +3\tau(2-\tau)f(\tau); \nonumber \\
A_{1/2}(\tau)&=&-2\tau [ 1+ (1-\tau)f(\tau)];\nonumber\\
\tilde{A}_{1/2}(\tau)&=&\tau f(\tau) ; \nonumber\\
A_0(\tau)&=&[1-\tau f(\tau)]
\end{eqnarray}
$\tau_i=\frac{4m_i^2}{m_{h_1}^2}$ for $i=t,W,H^{\pm}$ and 
\begin{equation}
f(\tau)=\arcsin^2(\frac{1}{\sqrt{\tau}}).
\end{equation}

Meanwhile for the $h_1\rightarrow gg$  decay  
\begin{equation}
\Gamma(h_1\rightarrow gg)=\frac{G_F\alpha^2_s m_{h_1}^3}{64\sqrt{2}\pi^3}\left[ \left|  a_t A_{1/2}(\tau_t) \right|^2 
+\left| 2 \tilde{a}_t \tilde{A}_{1/2}(\tau_t) \right|^2\right].
\label{h1gg}
\end{equation}

On the other hand, the the width of $h_1\rightarrow \gamma Z$ decay  is~\cite{Spira:1991tj,Spira:1997dg,GonzalezSprinberg:2004bb,Cordero-Cid:2013sxa}
\begin{eqnarray}
\label{h1gaz}
\Gamma(h_1\rightarrow \gamma Z)&=&\frac{G_F^2\alpha m_W^2 m_{h_1}^3}{64\pi^4}\left(1-\frac{m_Z^2}{m_{h_1}^2}\right)^3\nonumber \\
&&[ |a_t B_{1/2}(\tau_t,\lambda_t)+a_WB_1(\tau_W,\lambda_W) +B_0(\tau_{H^\pm},\lambda_{H^\pm})|^2+ | \tilde{a}_t \tilde{B}_{1/2}(\tau_t,\lambda_t) |^2],
\end{eqnarray}
where
\begin{eqnarray}
B_{1/2}(\tau,\lambda)&=&\frac{4(1-\frac{4}{3})}{\cos\theta_W}[I_1(\tau,\lambda)-I_2(\tau,\lambda)] \nonumber \\
\tilde{B}_{1/2}(\tau,\lambda)&=&\frac{4(1-\frac{4}{3})}{\cos\theta_W}I_2(\tau,\lambda) \nonumber \\
B_{1}(\tau,\lambda)&=&\cos\theta_W\left\{4(3-\tan^2\theta_W)I_2(\tau,\lambda)\right.\nonumber \\
&+&\left. [ (1+\frac{2}{\tau} )\tan^2\theta_W-(5+\frac{2}{\tau} ) ]I_1(\tau,\lambda) \right\} \nonumber \\
B_0(\tau,\lambda)  &=&\frac{\cos2\theta_W}{\cos\theta_W}I_1(\tau,\lambda)   
\end{eqnarray}
and
\begin{eqnarray}
I_1(\tau,\lambda)&=&\frac{\tau\lambda}{2(\tau-\lambda)}+\frac{\tau^2\lambda^2}{2(\tau-\lambda)^2}[f(\tau)-f(\lambda)] \nonumber \\
&&+\frac{\tau^2\lambda}{(\tau-\lambda)^2}[g(\tau)-g(\lambda)] \nonumber\\
I_2(\tau,\lambda)&=&-\frac{\tau\lambda}{2(\tau-\lambda)}[f(\tau)-f(\lambda)],\nonumber\\
\end{eqnarray}
with
\begin{equation}
g(\tau)=\sqrt{\tau-1}\arcsin^2\frac{1}{\sqrt{\tau}},
\end{equation}
$\tau_i=\frac{4m_i^2}{m_{W}^2}$ for $i=t,W,H^{\pm}$. In the Eqs. (\ref{h1gaga}), (\ref{h1gg}) and (\ref{h1gaz}), there is a term proportional to $\tilde{a}_t$ which represents the pseudoscalar coupling of the $h_1$ with $t\bar{t}$ at the loops. $\tilde{a}_t$ is proportional to $R_{13}=\sin\alpha_2$ which vanishes when the model is CP-Conserving.

For the $h_1\rightarrow WW^*$, $ZZ^*$ decays we will use the expressions reported in the literature~\cite{Keung:1984hn}, however, these expressions must be multiplied by additional factors denoted by $a_W^2$ or $a_Z^2$ and they are given in Eq. (\ref{coefficients-a}) respectively, which arise from the 2HDM-III. Note that if the matrix elements $R_{11}$, $R_{12}$ and $R_{13}$ are independent of the mixing parameter $\alpha_3$, see Eq. (\ref{r_matrix}),  then all decays for $h_1$ do not depend on it. The CPV effects are only a function of $\alpha_1$ and $\alpha_2$.

In order to obtain the branching ratios for $h_1$, we calculate the partial widths for $b\bar{b}$, $W W^*$, $Z Z^*$, $c \bar{c}$, $\tau \bar{\tau}$, $\mu \bar{\mu}$  at tree level, meanwhile $ g g$, $\gamma \gamma$, $\gamma Z$ at one loop in the 2HDM type III. We note that the $h_1$ does not have well-defined CP-parity and it couples to fermions with scalar and pseudoscalar interactions which contribute to partial width due to the Lagrangian Eq. (\ref{lagrangianIII}). Similarly, the decays at one loop level with top quark as internal line in the loop will also have two contributions arising from scalar and pseudoscalar couplings. If the mixing angle $\alpha_2=0$, the pseudoscalar couplings vanish and the partial decay widths are reduced to the 2HDM with CP conserving case. The radiative corrections due to QCD, QED and EW are considered in the numerical analysis from references~\cite{deFlorian:2016spz,Dabelstein:1991ky}. The masses of the fermions are also running to the scale of the SM boson mass $\sim125$ GeV.

\begin{figure}[htbp]
\centering 
\includegraphics[scale=0.7]{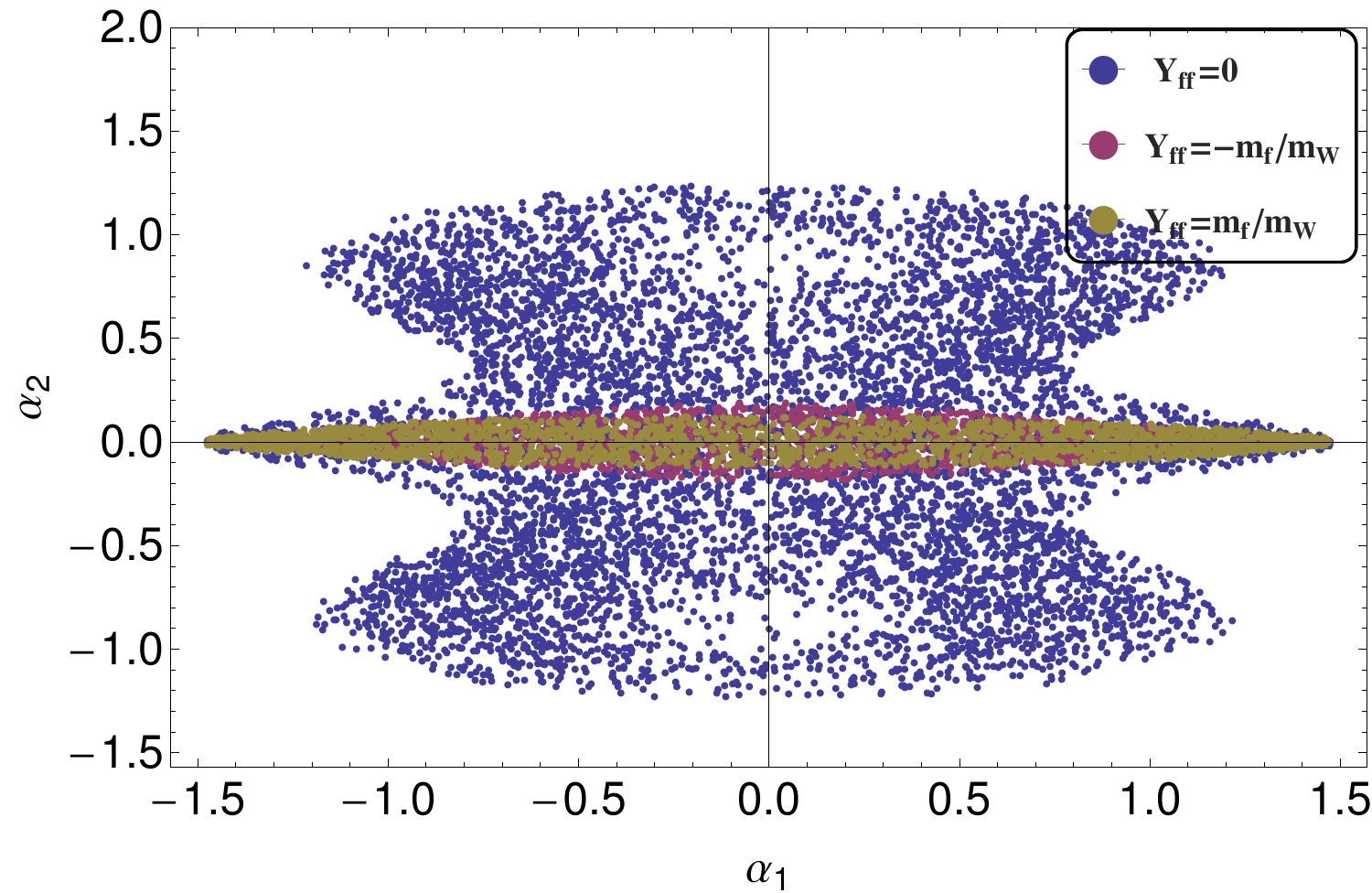}
\caption{\label{figurey0} Allowed regions for $\alpha_1$ and $\alpha_2$ with $Y_{ff}=0$ and $Y_{ff}=\pm\frac{m_f}{m_W}$, for $ f = e, \mu, \tau, b, c, s, d$, obtained through $\chi^2$ at 90$\%$ C.L. in $R_{XX}$.}
\end{figure}

\begin{figure}[htbp]
\centering 
\includegraphics[scale=0.7]{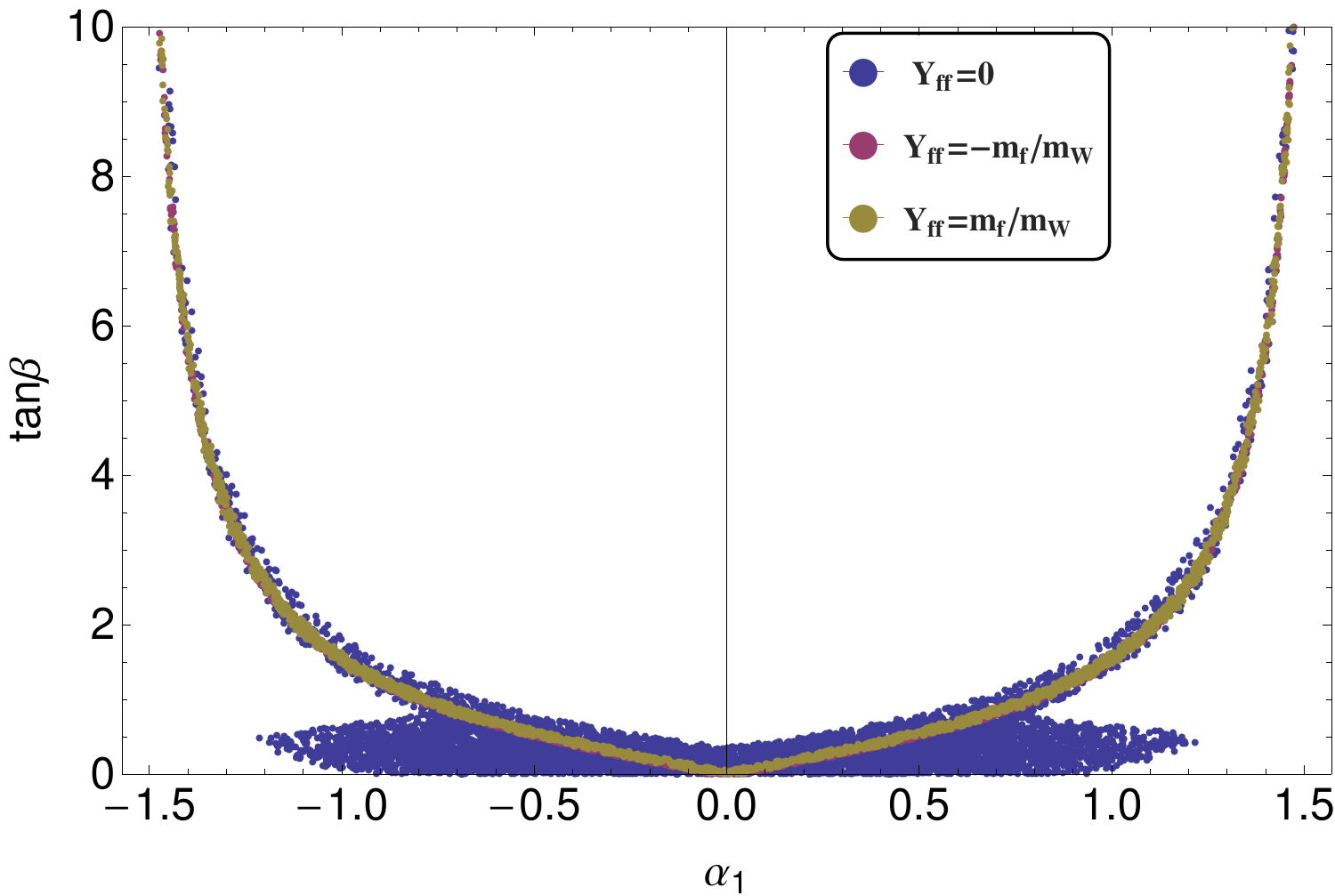}
\caption{\label{figurexia1} Allowed regions for $\alpha_1$ and $\tan\beta$ with $Y_{ff}=0$ and $Y_{ff}=\pm\frac{m_f}{m_W}$, for $ f = e, \mu, \tau, b, c, s, d$, obtained through $\chi^2$ at 90$\%$ C.L. in $R_{XX}$.}
\end{figure}

\begin{figure}[htbp]
\centering 
\includegraphics[scale=0.7]{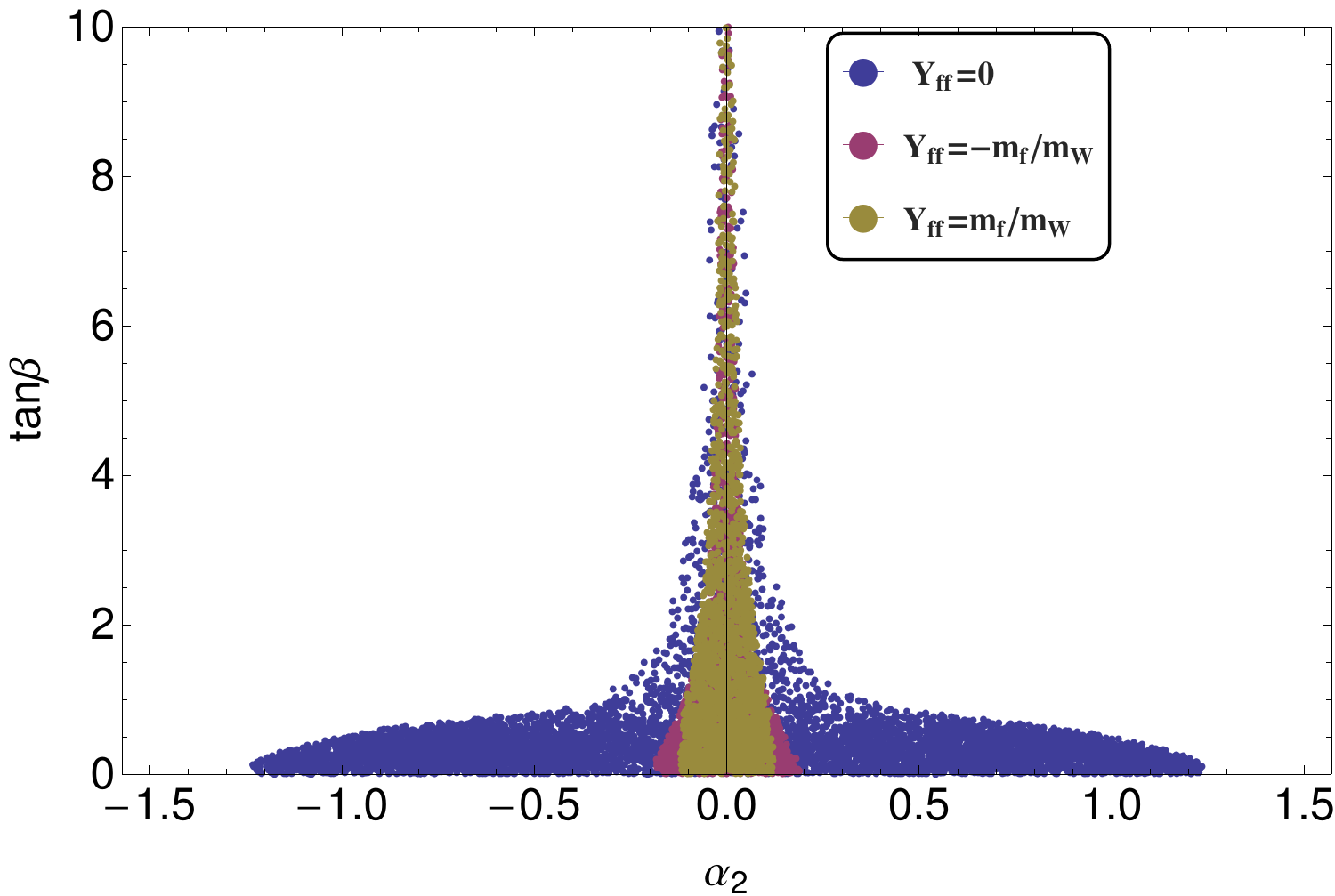}
\caption{\label{figurexia2} Allowed regions for $\alpha_2$ and $\alpha_2$ with $Y_{ff}=0$ and $Y_{ff}=\pm\frac{m_f}{m_W}$, for $ f = e, \mu, \tau, b, c, s, d$, obtained through $\chi^2$ at 90$\%$ C.L. in $R_{XX}$.}
\end{figure}

Following the measured at LHC of physical observables $R_{bb}$, $R_{WW}$, $R_{ZZ}$, $R_{\tau\tau}$ and $R_{\gamma\gamma}$  channels, we do an statistical analysis using  a $\chi^2$ function on the $\alpha_1$, $\alpha_2$ and $\tan\beta$ parameters for $m_{H^\pm} \approx 500$ GeV. To implement the $\chi^2$ analysis we take into account the reported observables by ATLAS and CMS for $R_{XX}$ shown in table \ref{rxx}. Figures \ref{figurey0}, \ref{figurexia1} and \ref{figurexia2} show the allowed values at $90\%$ C. L. for the mixing parameters assuming fixed values of the Yukawa couplings.

The model with $\tan\beta=0$ corresponds to one VEV equal to zero and its doublet is inert. However in this case we have CPV couplings with fermion. For $\alpha_2=0$ the allowed regions for $\alpha_1$ is $|\alpha_1|\lesssim 1.2$. We can see from Figure \ref{figurey0} that the allowed region is consistent with $\alpha_1=\alpha_2=0$ which corresponds to SM.
%
%
%
\begin{center}
\begin{table}
\caption{Reported values for $R_{XX}$ and $BR(H^0\rightarrow XX)$~\cite{Patrignani:2016xqp,deFlorian:2016spz}.}
  \begin{tabular}{ | c | c | c | }
    \hline
    $XX$ & $R_{XX}^{exp}$ & $BR(H^0\rightarrow XX)$ \\ \hline
    $b\, \bar{b}$ & $0.82\pm 0.30$ & $5.84\times 10^{-1}{}_{-3.3\%}^{+3.2\%}$ \\ \hline
    $W\, W^*$ & $1.08{}_{-0.16}^{+0.18}$ & $2.14\times 10^{-1}{}_{-4.2\%}^{+4.3\%}$ \\ \hline
    $g\, g$ & - & $7.3\times 10^{-2}{}_{-3.69\%}^{+3.61\%}$ \\ \hline
    $Z\, Z^*$ & $1.29{}_{-0.23}^{+0.26}$ & $2.62\times 10^{-2}{}_{-4.1\%}^{+4.3\%}$ \\ \hline
    $\tau\, \bar{\tau}$ & $1.12\pm 0.23$ & $6.27\times 10^{-2}{}_{-5.7\%}^{+5.7\%}$ \\ \hline
    $\gamma\, \gamma$ & $1.16\pm 0.18$ & $2.27\times 10^{-3}{}_{-4.9\%}^{+5\%}$ \\ \hline
    $c\, \bar{c}$ & - &  $5.1\times 10^{-3}{}_{-5.5\%}^{+5.5\%}$ \\ \hline
    $\gamma\, Z$ & - & $1.53\times 10^{-3}{}_{-8.9\%}^{+9\%}$ \\ \hline
    $\mu\, \bar{\mu}$ & -  & $2.18\times 10^{-4}{}_{-5.9\%}^{+6\%}$\\ \hline
  \end{tabular}
   \label{rxx}
  \end{table}
\end{center}
%

There is another interesting scenario in this model when $Y_{ff}\neq0$. The Yukawa couplings $Y_{ff}$ are restricted to be smaller than one in order to have a perturbative theory. These Yukawa couplings can be parametrized by the assumption of certain structure or Ansatz which is motivated when some elements of the Yukawa matrix in the interaction basis are fixed to zero~\cite{Du:1992iy,Hall:1993ni,Fritzsch:1994yx,Fritzsch:1995nx,Fritzsch:1997fw,Fritzsch:1997st,Fritzsch:1999rb,Branco:1999nb,Rosenfeld:2001sc,Chkareuli:2001dq,Fritzsch:2002ga,Matsuda:2006xa,Koide:2002cj,Matsuda:2003zm,CarcamoHernandez:2005ka,Crivellin:2013wna}. We consider for Yukawa couplings as limit values the Cheng-Sher Ansatz, $Y_{f_if_j}\approx \frac{\sqrt{m_{f_i}m_{f_j}}}{m_W}$~\cite{Cheng:1987rs}. Figure \ref{figurey0} shows the allowed regions for $\alpha_1$ and $\alpha_2$ with Yukawa couplings fixed in the extreme values, $Y_{ff}=\frac{m_f}{m_W}, -\frac{m_f}{m_W}$ and $Y_{ff}=0$. The last case, $Y_{ff}=0$, provides the greatest allowed region for mixing parameters.

The elements of the Yukawa matrices in Eq. (\ref{yukawa_quarks}) can be complex parameters in the most general case. However, the imaginary part of these couplings is strongly restricted by the Electric Dipole Moment (EDM) of the neutron~\cite{Brod:2013cka}. In particular, restrictions over imaginary part of the Yukawa couplings are obtained in the 2HDM with CP violation ~\cite{Gorbahn:2014sha}. Nevertheless, in this work the Yukawa couplings are assumed to be real parameters and $Y_{i j} = Y_{j i}$.

%
%
%
%
%

In Figure \ref{figurexia1} and \ref{figurexia2} we show regions in the plane $\alpha_1$-$\tan\beta$ and $\alpha_2$ -$\tan\beta$, respectively, for fixed values of Yukawa couplings, meanwhile Figure \ref{figurexia2} is for $\alpha_2$ and $\tan\beta$. Note that  $\alpha_2=0$ is a particular case of 2HDM without CPV. In this case Figure \ref{figurexia1} shows the values $\tan\beta>0$  and $|\alpha_1|\lesssim 1.5$ are preferred. On the other side, when $\alpha_2\neq 0$ the allowed region is reduced, for instance, in the case of $|\alpha_2 | \gtrsim 0.2$ the allowed region is bounded by $\tan\beta<2$ as shown in Figure \ref{figurexia2}.

%
\section{Rare top decays}
\label{sec4}
The observation of rare top decays with FCNC would be as a clear signal of physics beyond SM which can be understood in extended model.
We will analyze the rare top decays $t\rightarrow cZ$ and $t\rightarrow c H^0$ in this section, while the $t\rightarrow c\gamma$ and $t\rightarrow cg$ have been previously studied~\cite{Gaitan:2015hga}. In the 2HDM type III, the neutral scalar field has a scalar and pseudoscalar coupling with the top quark which contributes inside the loop associated with the Higgs decay into two photons.
%
%
\begin{figure*}[!h]
\centerline{
(a)\includegraphics[width=2.5in]{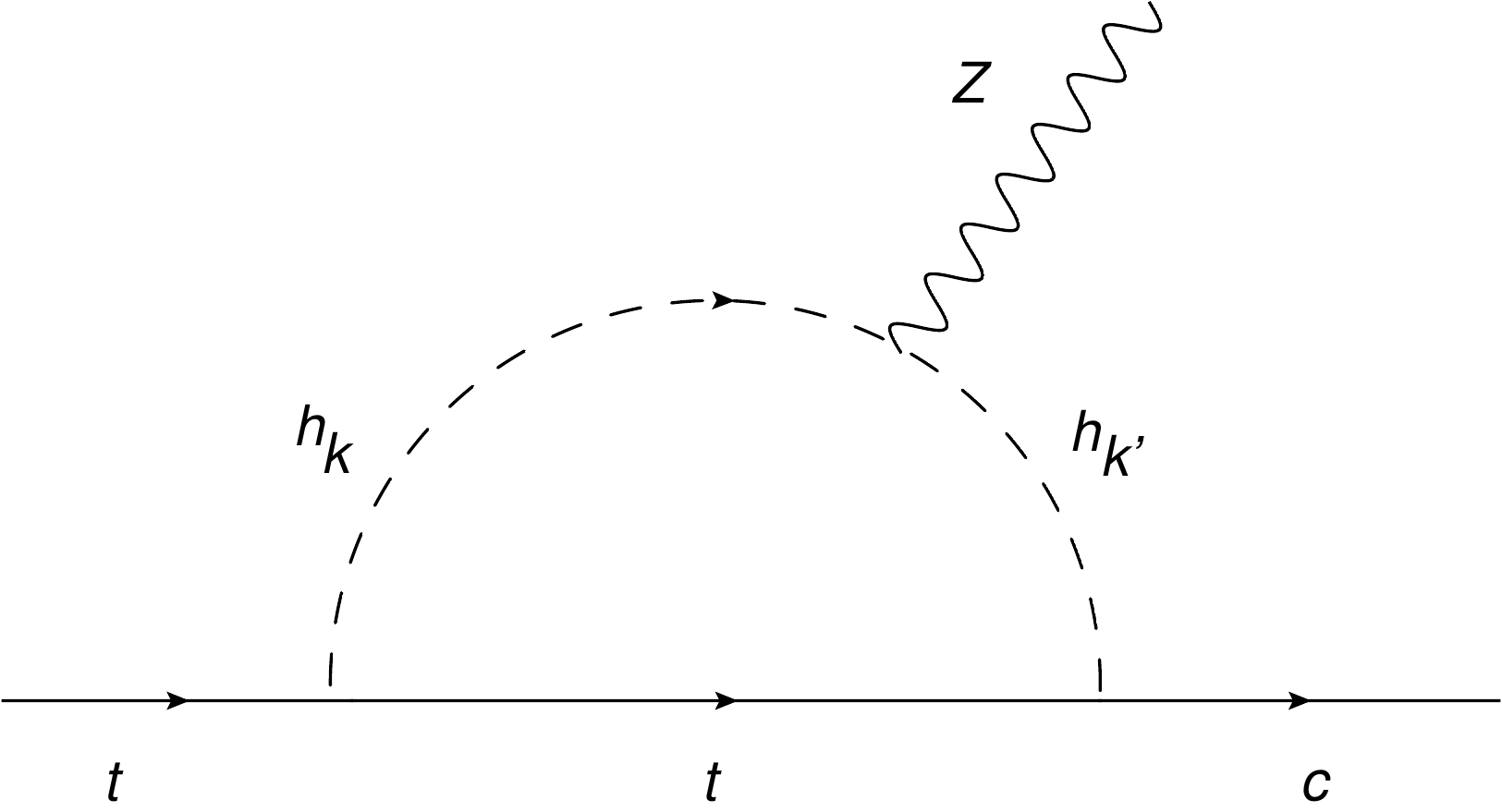}
\label{fig:fig1_li-lj}
\hfil
(b)\includegraphics[width=2.5in]{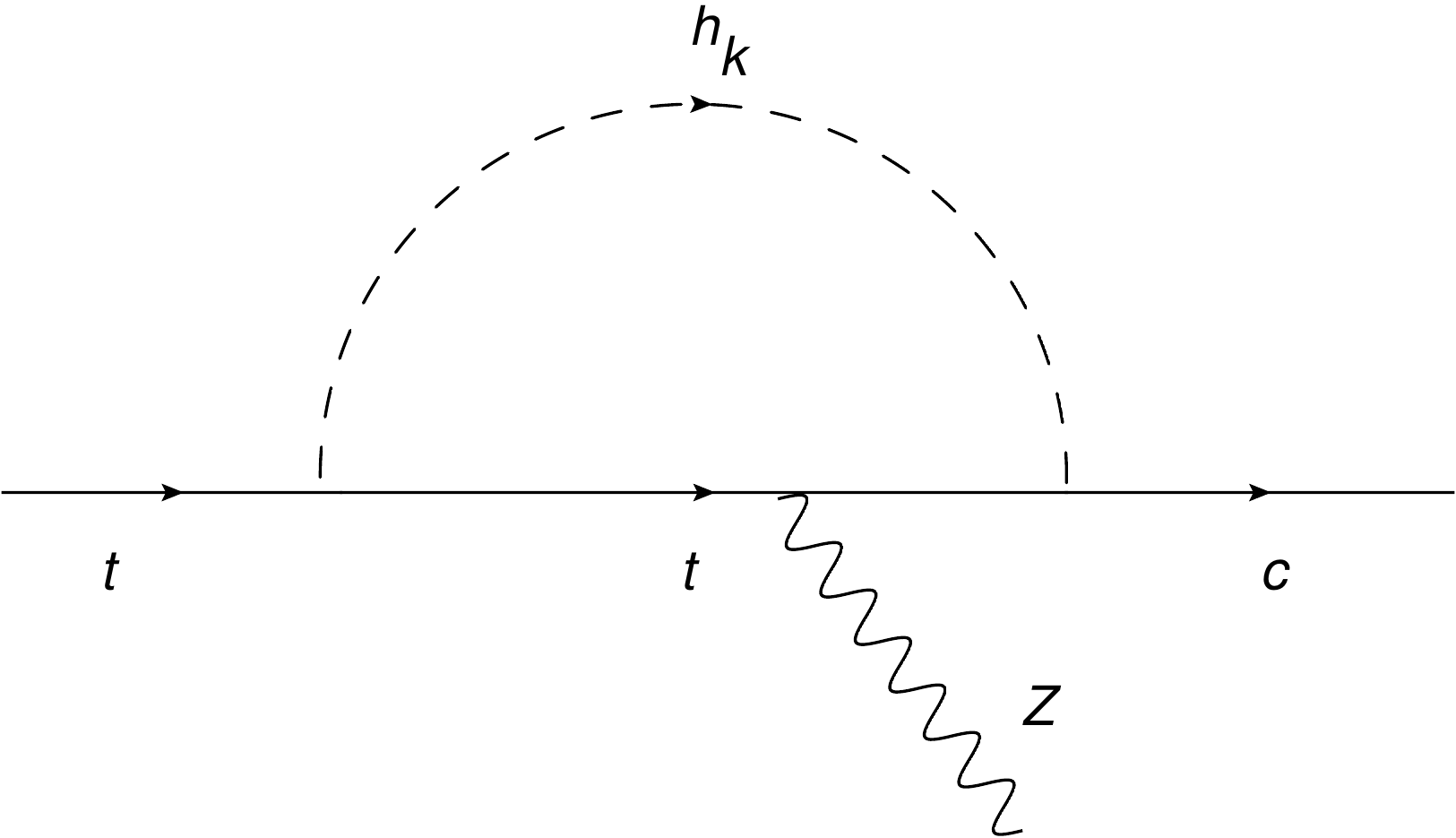}%
\label{fig:fig2_li-lj}
}
\hfil
(c)\includegraphics[width=2.5in]{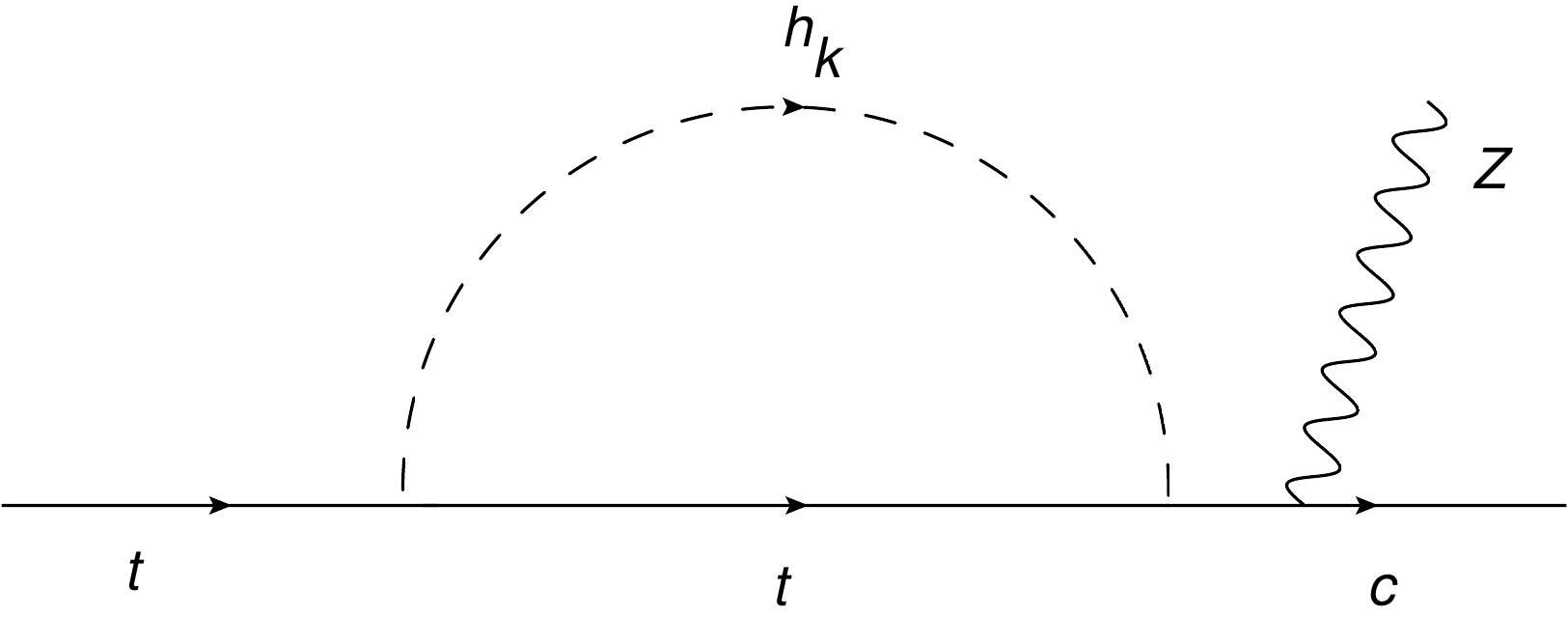}
\label{fig:fig1_li-lj}
\hfil
(d)\includegraphics[width=2.5in]{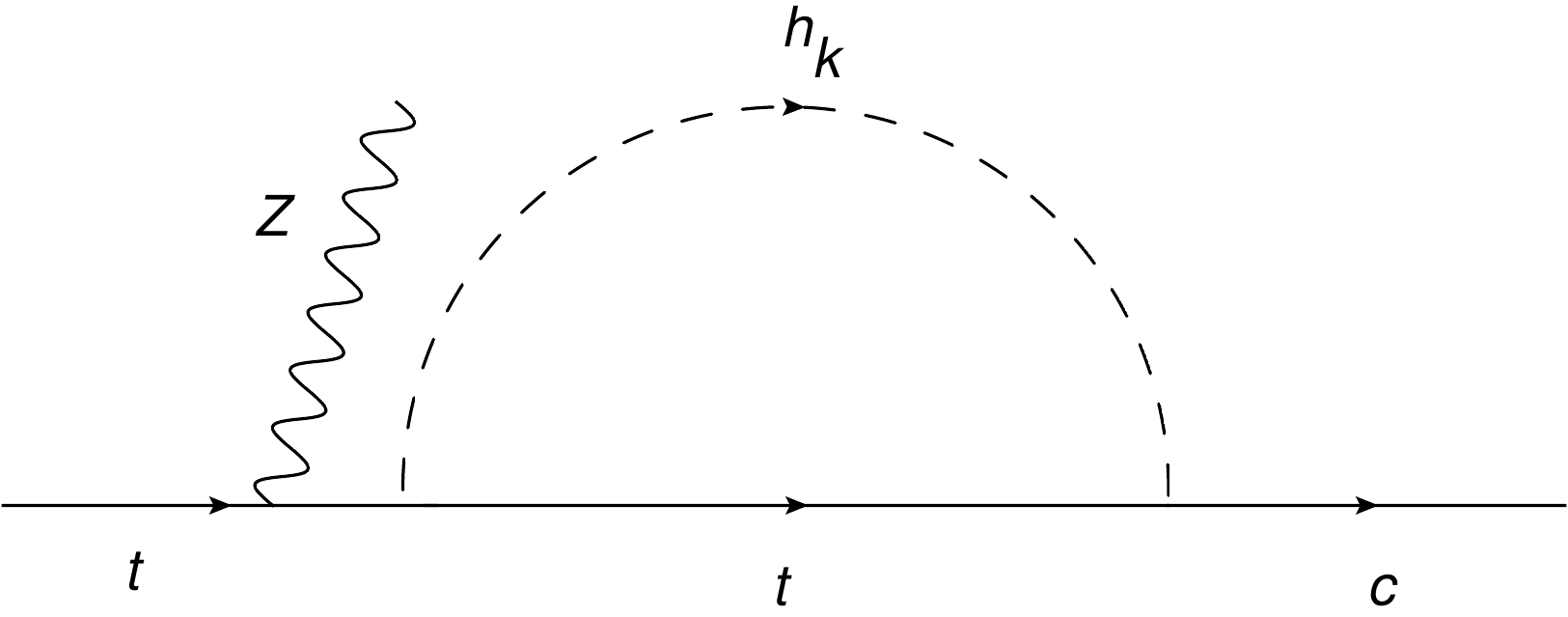}%
\label{fig:fig2_li-lj}
\caption{Feynman diagrams for the $t\rightarrow cZ$ decay.}
\label{feynman}
\end{figure*}
%
%

\subsection{$t\rightarrow c Z$}
The contributions from neutral scalars with FC to the amplitude for $t\rightarrow c \,Z  $ decay are shown in figure \ref{feynman}. In general, the amplitude associated to the Feynman diagrams in figure \ref{feynman} is written as
\begin{equation}
\mathcal{M}(t\rightarrow c \,Z)=\bar{u}\left(c\right)\left[\gamma_\mu\left(V_LP_L+V_RP_R\right)+\frac{2i}{m_t}p_\mu\left(F_LP_L+F_RP_R\right) \right]u\left(t\right)\epsilon_Z^\mu
\label{ampG}
\end{equation}
Here the $q_\mu$ contributions are not considered due to the gauge condition $\epsilon_Z\cdot q=0$. The form factors associated with $\gamma_\mu$ and $p_\mu$ can be related through the Gordon identity, 
\begin{equation}
\label{gordon}
\frac{2ip_\mu}{m_t}P_{L,R}=\gamma_\mu P_{R,L}+\frac{i}{m_t}\sigma_{\mu\nu}q^{\nu}P_{L,R}.
\end{equation}
In the amplitude $\mathcal{M}(t\rightarrow c \,Z)$, the Gordon identity can be approximated as $\frac{2p_\mu}{m_t}P_{L,R}\approx \gamma_\mu P_{R,L}$ and the amplitude of $t\rightarrow cZ$ at one loop can be written as:
\begin{equation}
\label{amp}
\mathcal{M}(t\rightarrow c Z)=\frac{-i}{16\pi^2}\frac{g^2m_tY_{tc}}{M_W\cos\theta_W}\bar{u}\left(c\right)\sum_{D=a,b,c,d}\left\{\gamma_\mu\left[\left(\tilde{V}_L^D+\tilde{F}_R^D\right)P_L+\left(\tilde{V}_R^D+\tilde{F}_L^D\right)P_R\right]\right\}u\left(t\right)\epsilon_Z^\mu.
\end{equation}
where $D=a, b, c, d$.
The explicit values for all dimensionless form factors $\tilde{V}_{L,R}^{D}$ and $\tilde{F}_{L,R}^{D}$, after dimensional regularization, for Feynman diagrams  are written in the appendix. All contributions are finite because there is not $tcZ$ vertex at tree level and the Lagrangian can not be renormalized.  The amplitude in Eq. (\ref{amp}) is used to obtain   
\begin{equation}
\label{widthZ}
\Gamma(t\rightarrow c Z)=\frac{G_F^2 m_t^5 Y_{tc}^2}{256\pi^5}\left( 1- \hat{M}_Z^2\right)\left( 1+ 2\hat{M}_Z^2\right)\left( \mid \tilde{A}\mid^2+\mid \tilde{B}\mid^2\right),
\end{equation}
where $\hat{M_Z}=\frac{M_Z}{m_t}$. The dimensionless terms $\tilde{A}$ and $\tilde{B}$ are 
\begin{equation}
\label{Atilde}
\tilde{A}=\sum_{D=a,b,c,d}\left(\tilde{V}_L^D+\tilde{F}_R^D \right),
\end{equation}
\begin{equation}
\label{Btilde}
\tilde{B}=\sum_{D=a,b,c,d}\left(\tilde{V}_R^D+\tilde{F}_L^D \right)
\end{equation}
In order to obtain the branching ratio $BR(t\rightarrow cZ)$ the SM width for the top quark can be approximated to $tbW$ width as~\cite{Patrignani:2016xqp} 
\begin{equation}
\Gamma_{\textrm{top}}=\frac{G_f m_t^3}{8\pi\sqrt{2}} \left(1-\frac{M_W^2}{m_t^2}\right)^2 \left(1+2\frac{M_W^2}{m_t^2}\right)\left[1-\frac{2\alpha_s} {3\pi}  \left( \frac{2\pi^2}{3}-\frac{5}{2} \right)\right].
\end{equation}
In the 2HDM-III with CPV, we include the FCNSI contributions, Eq. (\ref{widthZ})  in the total width for top quark, such that, $\Gamma_{total}=\Gamma_{\textrm{top}}+\Gamma_{\textrm{rare decays}}$, where $\Gamma_{\textrm{rare decays}} = \Gamma(t\rightarrow c Z) +  \Gamma(t\rightarrow ch_1)$. The dominant contribution is the $\Gamma_{\textrm{top}}$; however, the $\Gamma_{\textrm{rare decays}}$ contribution, which contains $\Gamma(t\rightarrow ch_1)$ at tree level,  can reach up to $\sim1\%$ for specific values of the model parameters, as shows in the next subsection. When only the SM contribution is considered, the branching ratio for $t\rightarrow cZ$ can be approximated as
\begin{equation}
\label{BRZ}
BR(t\rightarrow cZ)\approx \frac{G_F^2 m_t^2 Y^2_{tc}}{ 16\sqrt{2}\pi^4}\frac{\left( 1- \hat{M}^2_Z\right)}{\left( 1- \hat{M}^2_W\right)}\frac{\left( 1+2\hat{M}^2_Z\right)}{\left( 1+2\hat{M}_W^2\right)}\left( \mid \tilde{A}\mid^2+\mid \tilde{B}\mid^2\right).
\end{equation}
The $\mid \tilde{A}\mid^2$ and $\mid \tilde{B}\mid^2$ are functions of the neutral scalar masses $m_{h_k}$, $k=1,2,3$, and of the mixing parameters $\alpha_{1,2,3}$, $\beta$. $h_1$ is the SM Higgs mass, $m_{h_1}=125$ GeV~\cite{Patrignani:2016xqp}, and we fix the masses of the neutral scalars $h_{2,3}$ of the order of  600 $GeV$. 
%
%
%
\begin{figure*}[!h]
\centerline{
(a)
\includegraphics[scale=0.5]{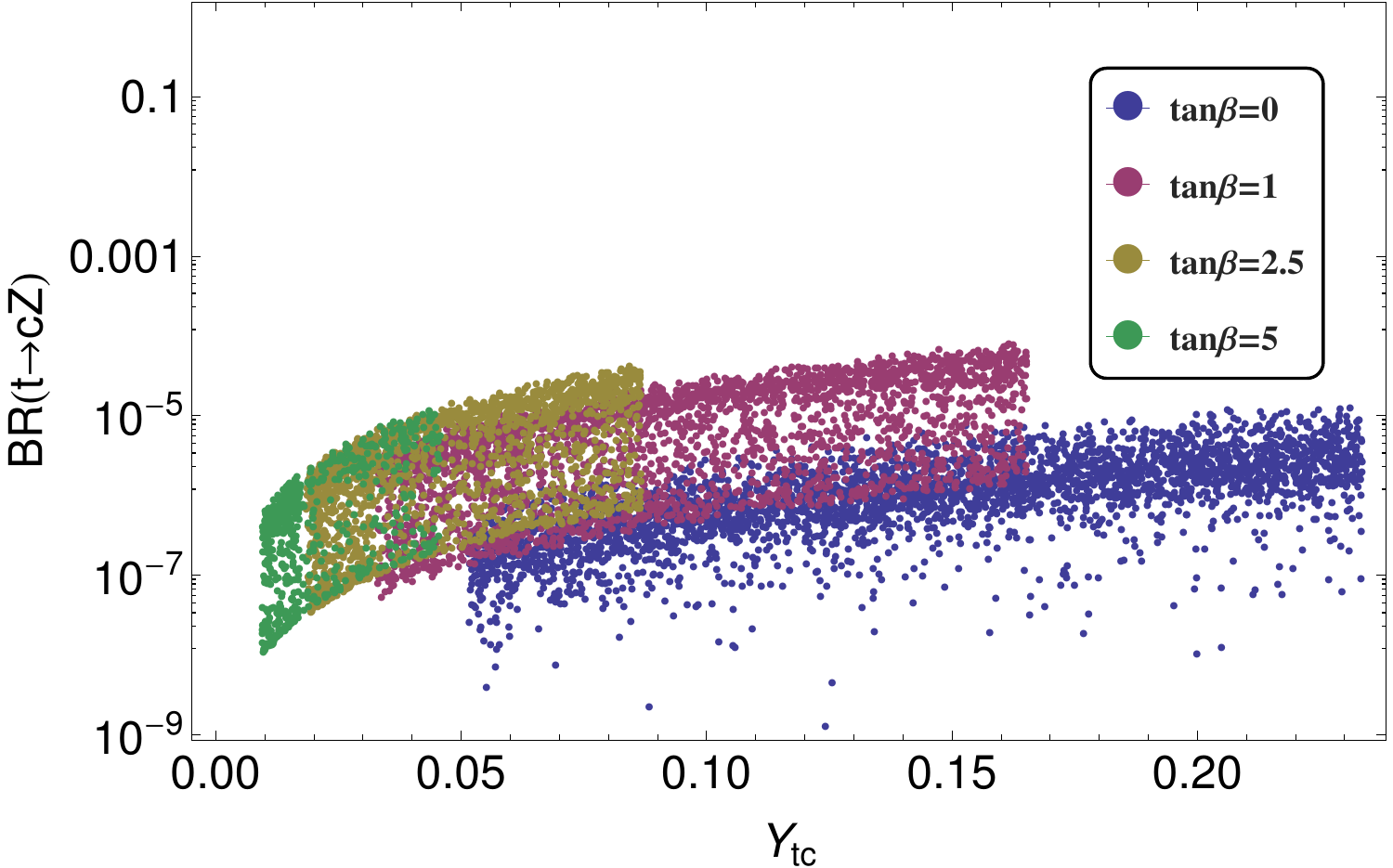}
\label{BRz_ytc_a}
\hfil
(b)
\includegraphics[scale=0.5]{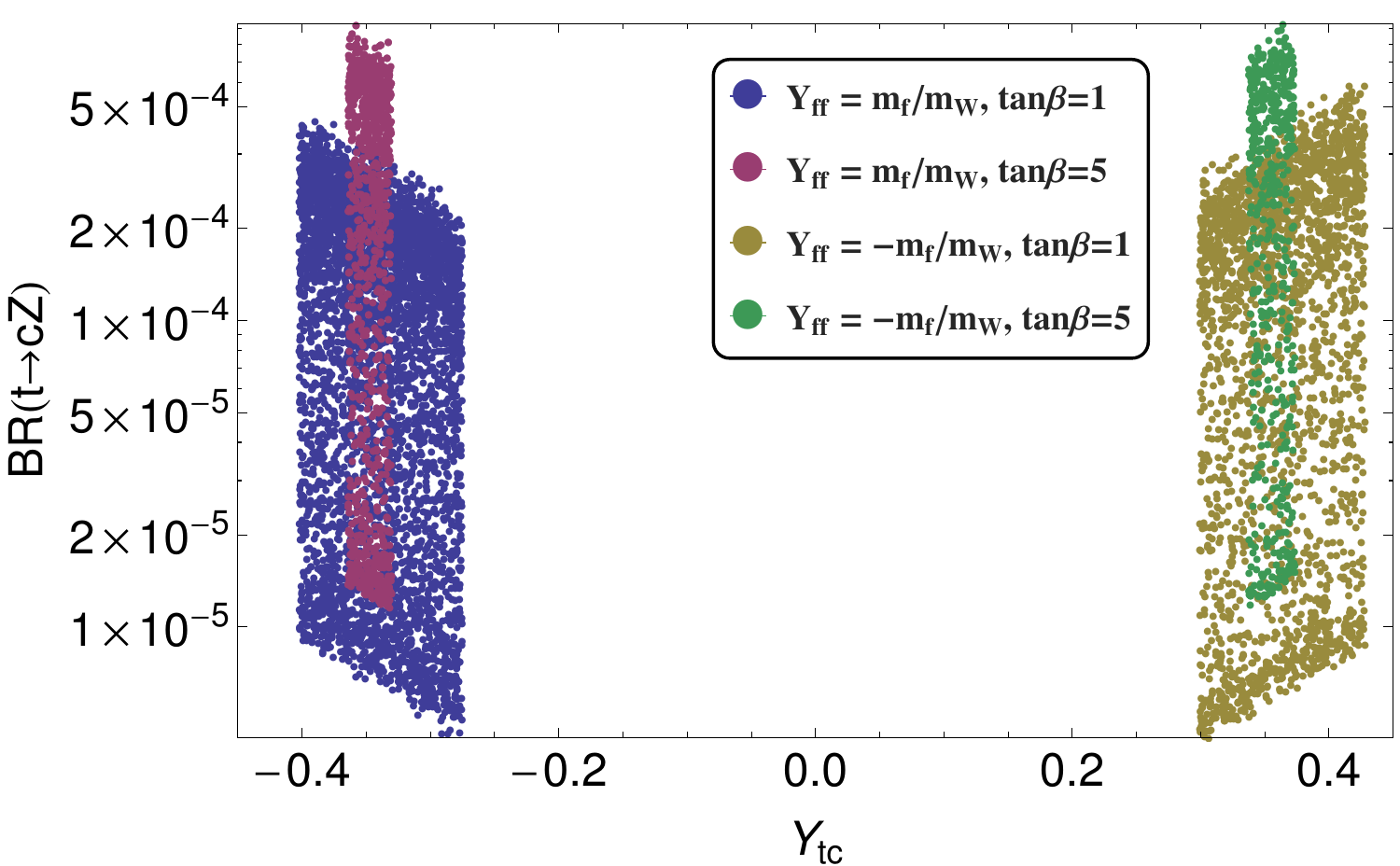}
\label{BRz_ytc_b}
}
\hfil
\caption{\label{BRz_ytc} $BR(t\rightarrow cZ)$ as function of $Y_{tc}$ in the allowed regions for $\alpha_{1,2}$. Figure shows the behavior for $Y_{ff}=\pm m_f/m_W$, for $ f = e, \mu, \tau, b, c, s, d$.}
\label{BRz_ytc}
\end{figure*}
%
%
%
\begin{figure*}[!h]
\centerline{
(a)\includegraphics[scale=0.5]{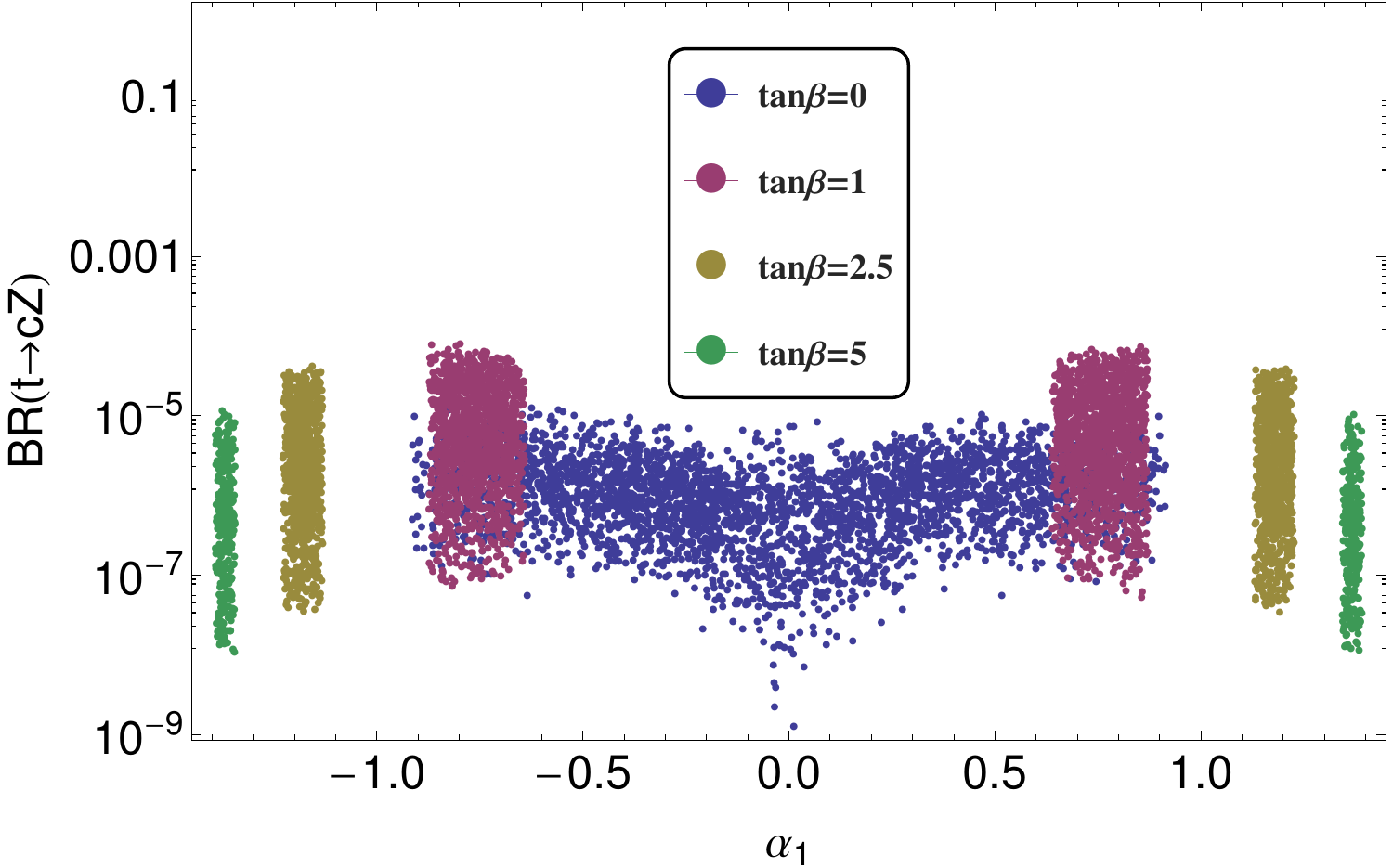}\label{BRz_a1_a}
\hfil
(b)\includegraphics[scale=0.5]{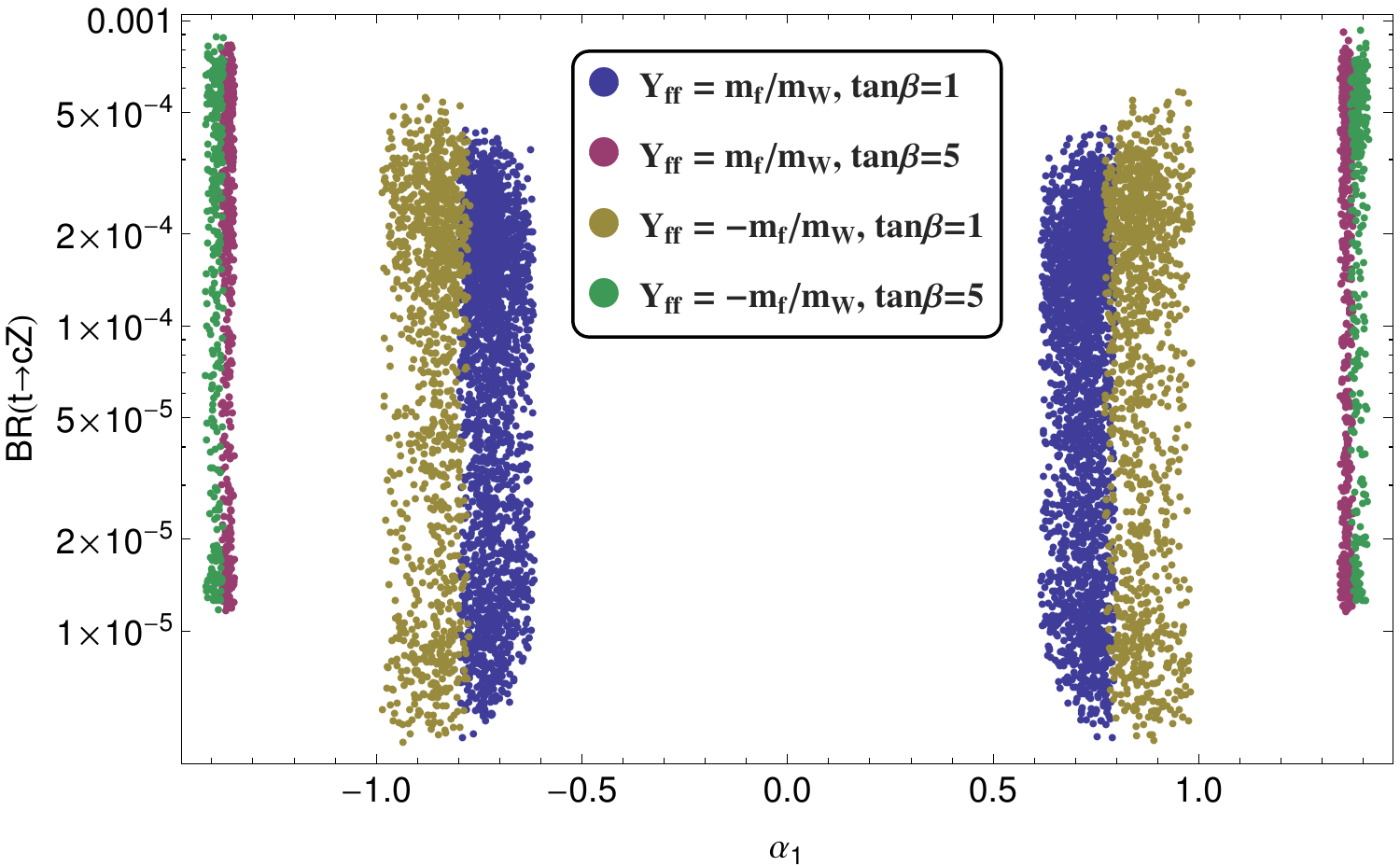}\label{BRz_a1_b}
}
\hfil
\caption{\label{BRz_a1} $BR(t\rightarrow cZ)$ as function of $\alpha_1$ for fixed values of $\tan\beta$ in the allowed regions for $\alpha_{1,2}$. Figure (a) for $Y_{ff}=0$, while figure (b) for $Y_{ff}=\pm m_f/m_W$, for $ f = e, \mu, \tau, b, c, s, d$.}
\label{BRz_a1}
\end{figure*}
%
%
%
\begin{figure*}[!h]
\centerline{
(a)\includegraphics[scale=0.5]{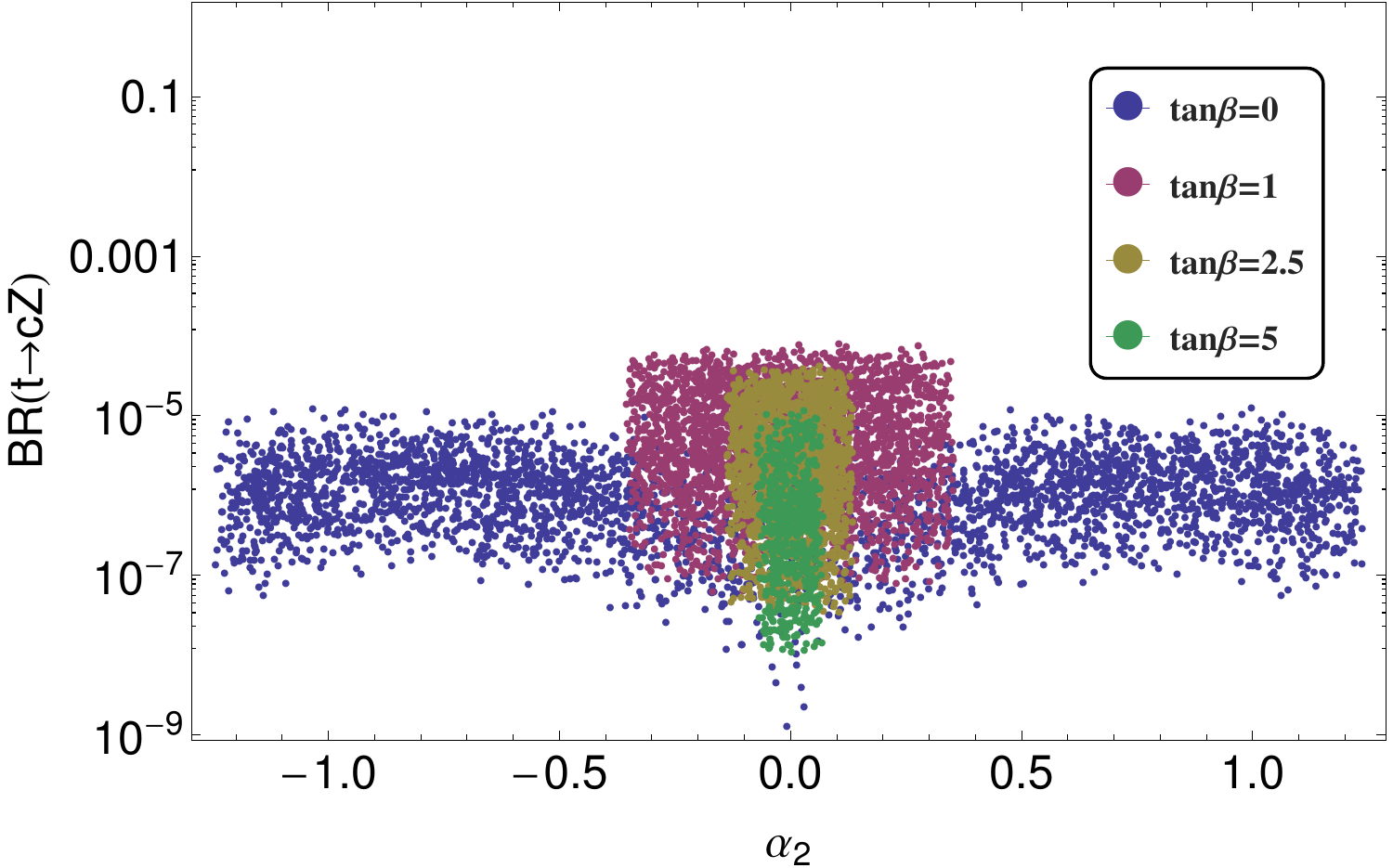}\label{BRz_a2_a}
\hfil
(b)\includegraphics[scale=0.5]{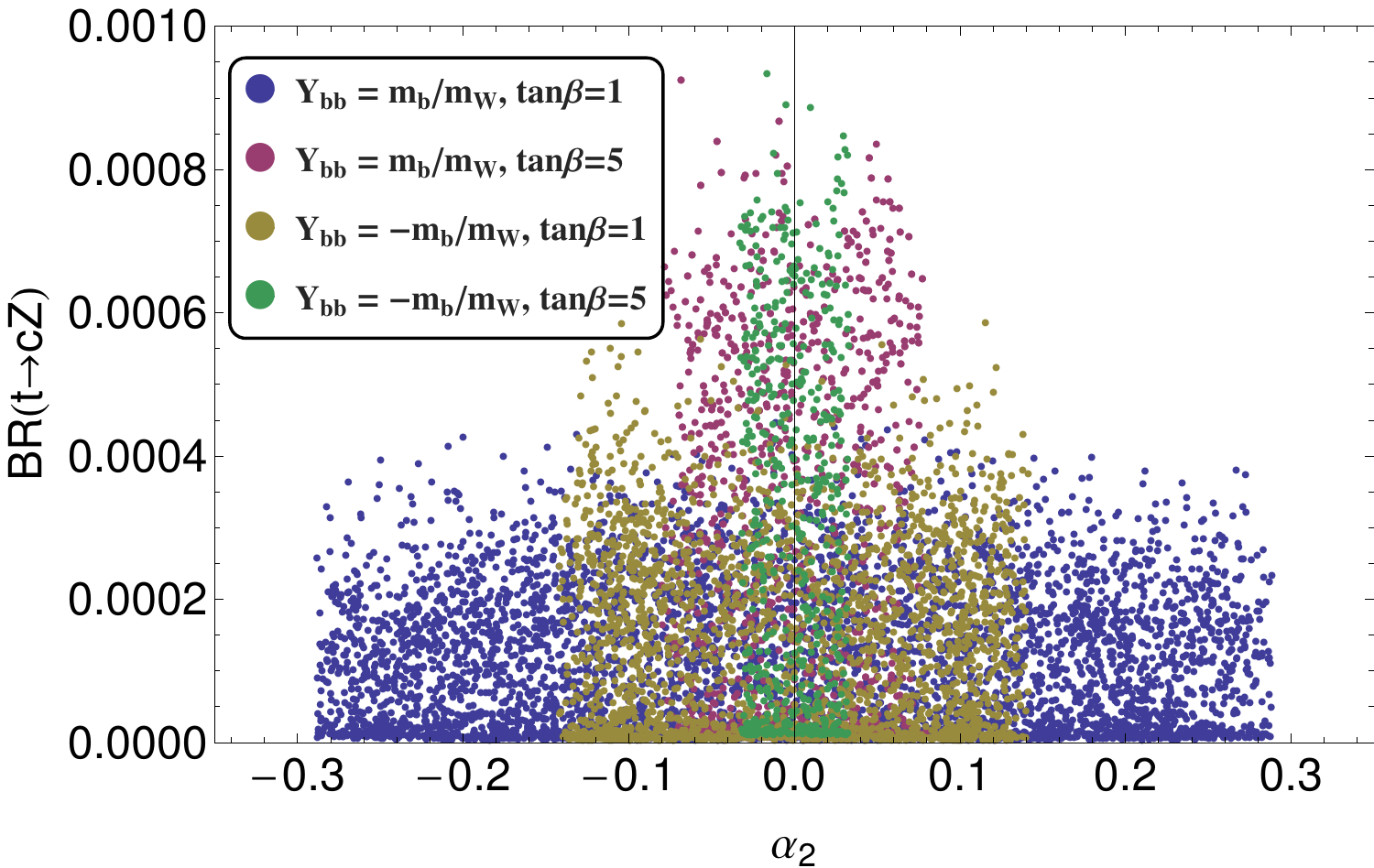}\label{BRz_a2_b}
}
\hfil
\caption{\label{BRz_a2} $BR(t\rightarrow cZ)$ as function of $\alpha_2$ for fixed values of $\tan\beta$ in the allowed regions for $\alpha_{1,2}$. Figure (a) for $Y_{ff}=0$, while figure (b) for $Y_{ff}=\pm m_f/m_W$, for $ f = e, \mu, \tau, b, c, s, d$.}
\label{BRz_a2}
\end{figure*}
%
%
%
%
\subsection{$t\rightarrow c H$}
\label{subtch}
In this subsection, we will analyze the FC top decay $t\rightarrow c h_{1}$ in this model which can occurs at tree level. The coupling for this decay is given in Eq. (\ref{yukawa_quarks}) and again the non vanishing  $Y_{tc}$ is responsible for the flavor change through the neutral scalar mediation. The partial decay width is 
\begin{equation}
\Gamma(t\rightarrow c h_1)= \frac{m_t}{8\pi}\left( 1-\hat{m}_{h_1}^2\right)| Y_{tc}|^2 | B_1|^2,
\label{tch}
\end{equation}
where 
\begin{equation}
B_1=\cos\alpha_2\cos\left( \alpha_1-\beta \right)+ i \sin\alpha_2
\label{B1}
\end{equation}
is obtained from Eq. (13). Figures \ref{BRh_ytc}, \ref{BRh_0a}  and \ref{BRh_a} show the behavior of this $BR (t\rightarrow c h_1)$ as function of the mixing parameters. 

%
%
%
\begin{figure*}[!h]
\centerline{
(a)
\includegraphics[scale=0.5]{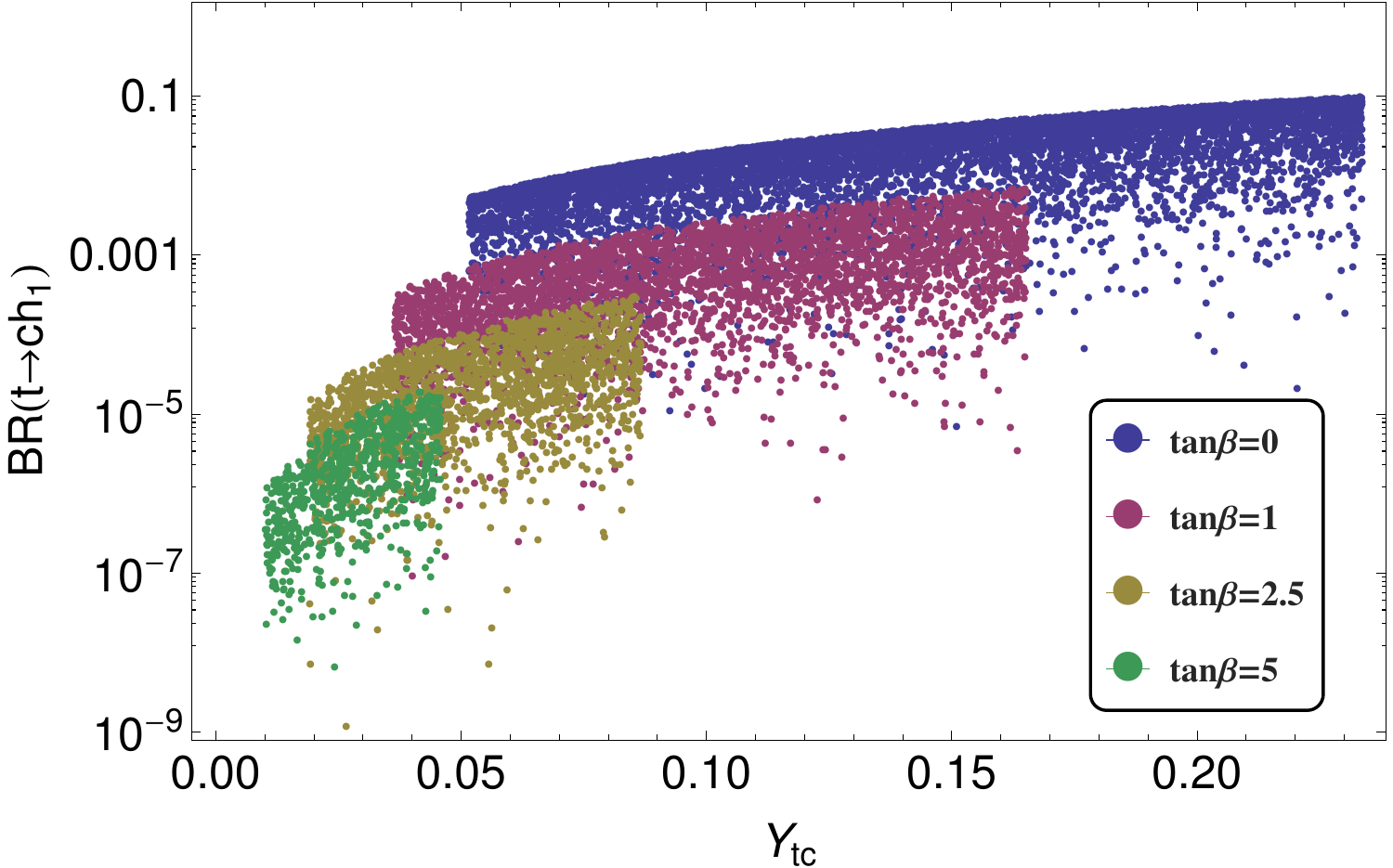}
\label{BRh_ytc_a}
\hfil
(b)
\includegraphics[scale=0.5]{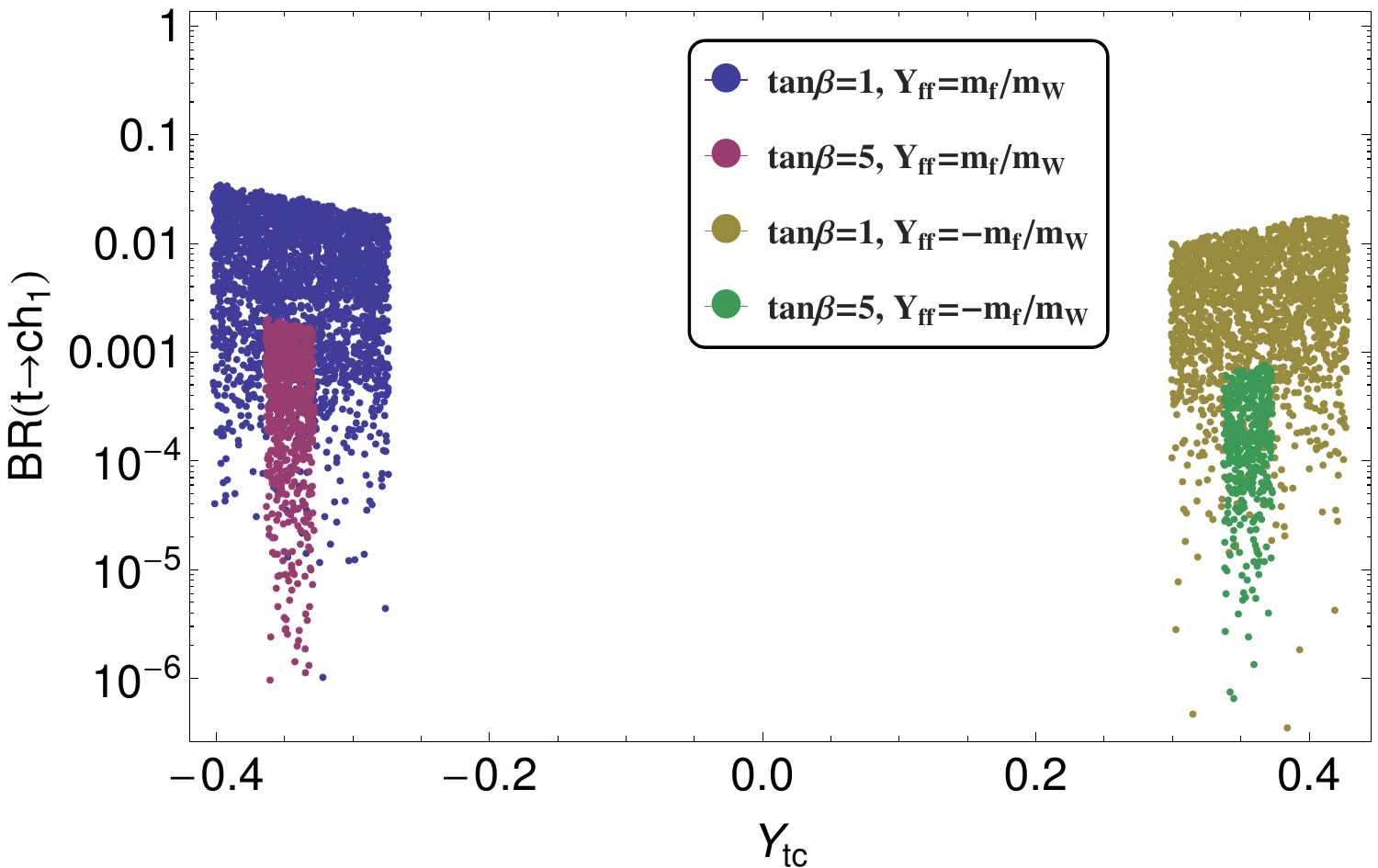}
\label{BRh_ytc_b}
}
\hfil
\caption{\label{BRh_ytc} $BR(t\rightarrow ch_1)$ as function of $Y_{tc}$ for fixed values of $\tan\beta$ in the allowed regions for $\alpha_{1,2}$. Figure (a) for $Y_{ff}=0$, while figure (b) for $Y_{ff}=\pm m_f/m_W$.}
\label{BRh_ytc}
\end{figure*}
%
%
%
%
\begin{figure*}[!h]
\centerline{
(a)
\includegraphics[scale=0.5]{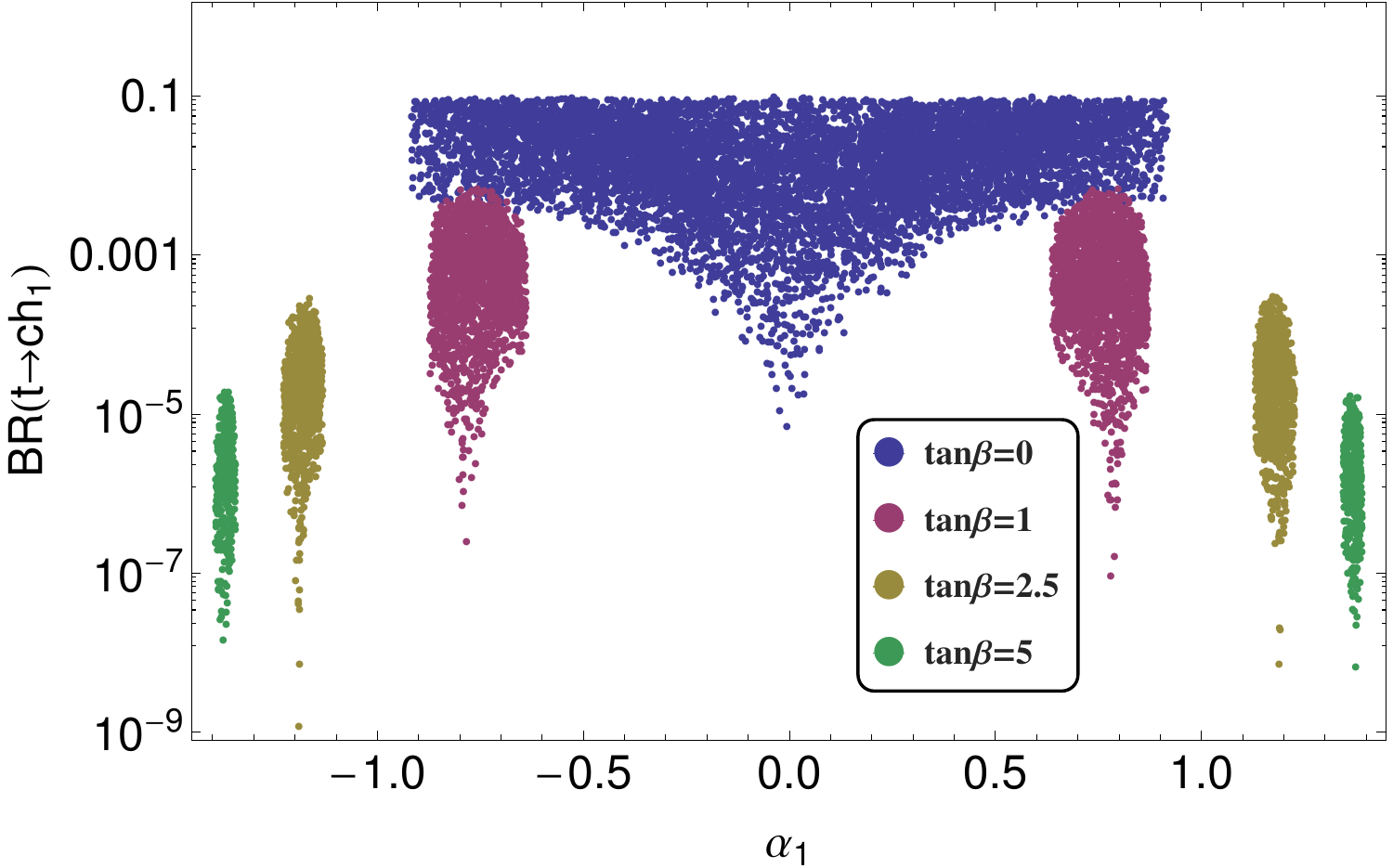}
\label{BRh_0a_a}
\hfil
(b)
\includegraphics[scale=0.5]{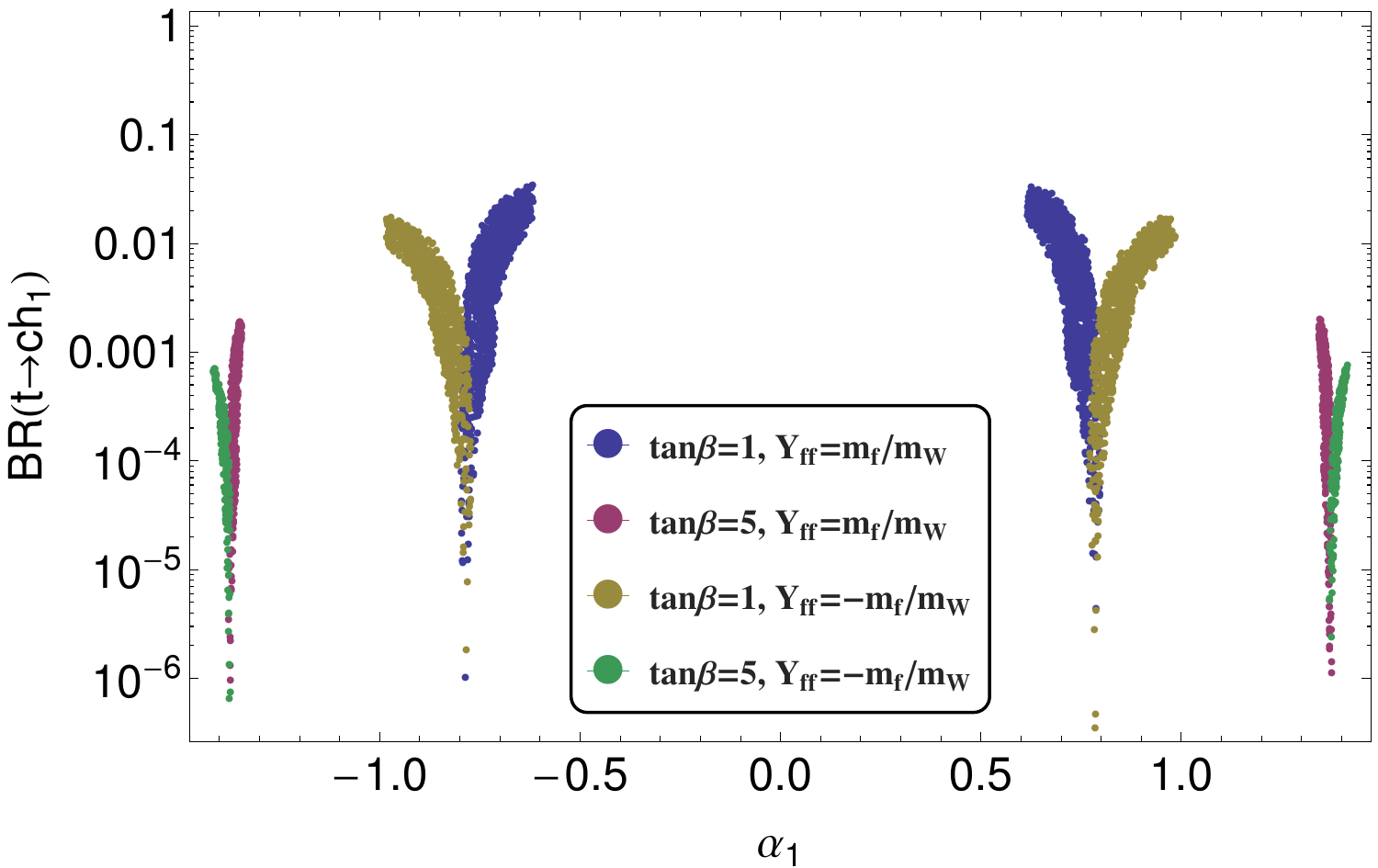}
\label{BRh_0a_b}
}
\hfil
\caption{\label{BRh_0a} $BR(t\rightarrow ch_1)$ as function of $\alpha_1$ for fixed values of $\tan\beta$ in the allowed regions for $\alpha_{1,2}$. Figure (a) for $Y_{ff}=0$, while figure (b) for $Y_{ff}=\pm m_f/m_W$.}
\label{BRh_0a}
\end{figure*}
%
%
%
\begin{figure*}[!h]
\centerline{
(a)
\includegraphics[scale=0.5]{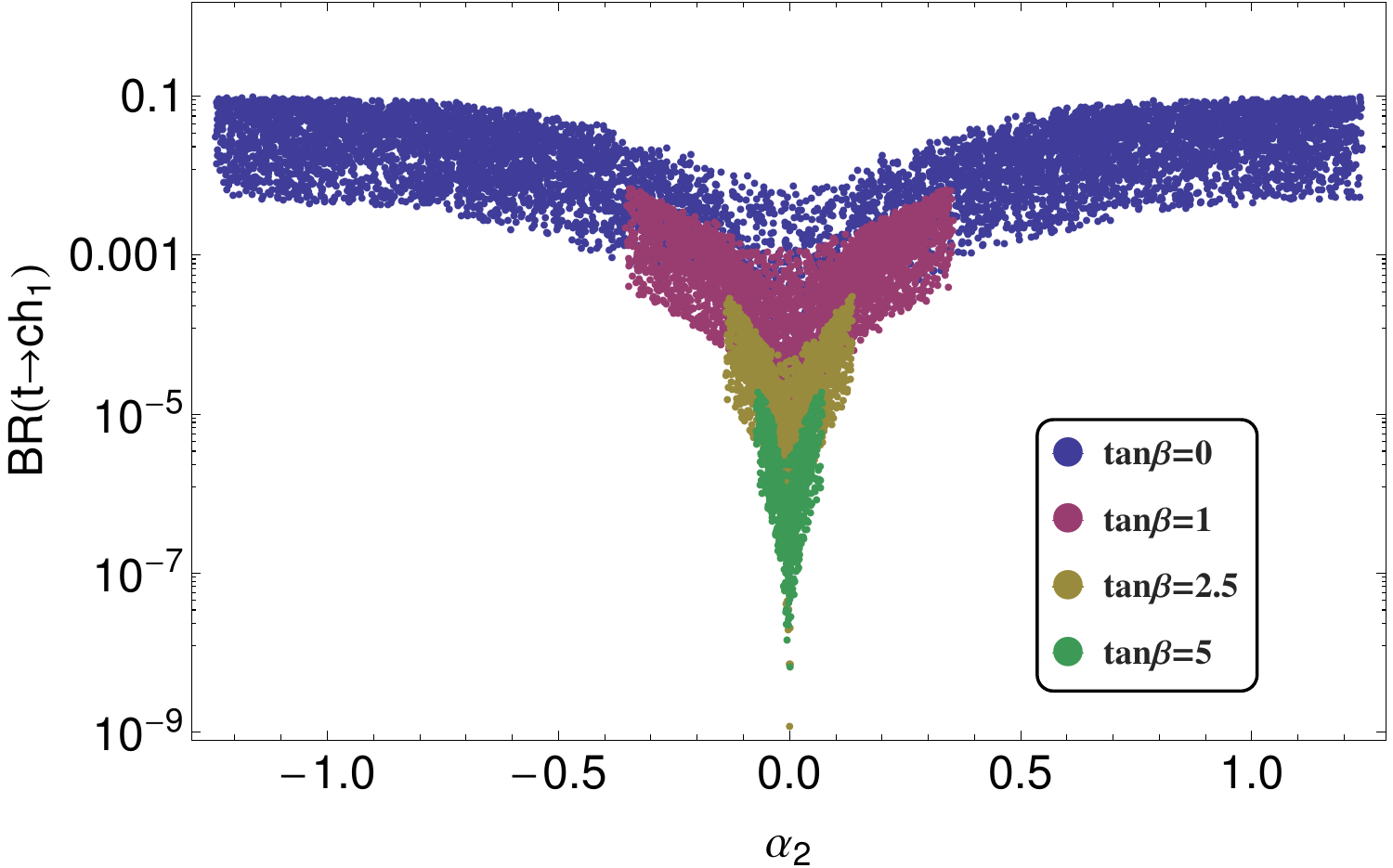}
\label{BRh_a_a}
\hfil
(b)
\includegraphics[scale=0.5]{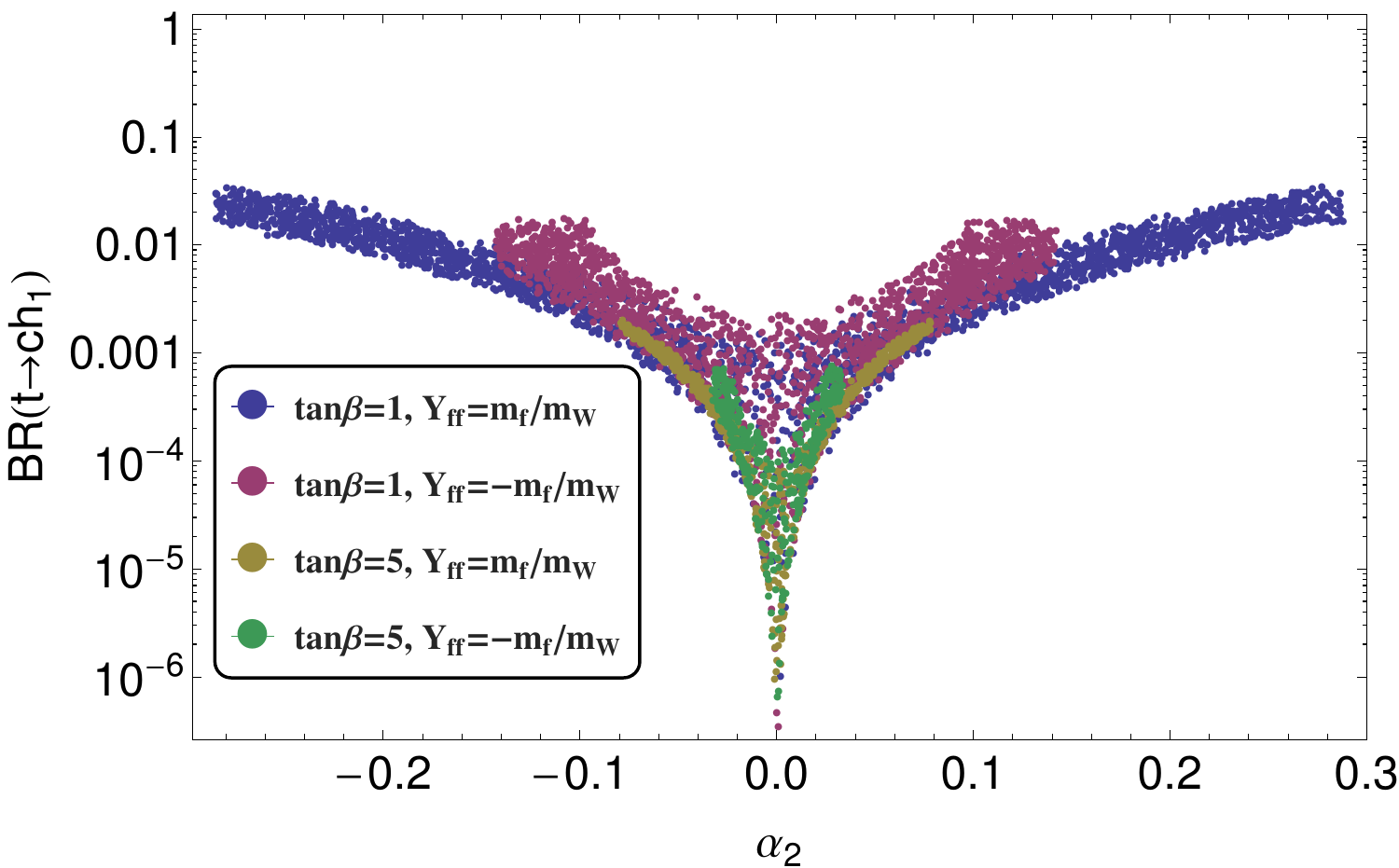}
\label{BRh_a_b}
}
\hfil
\caption{\label{BRh_a} $BR(t\rightarrow cZ)$ as function of $\alpha_2$ for fixed values of $\tan\beta$ in the allowed regions for $\alpha_{1,2}$. Figure (a) for $Y_{ff}=0$, while figure (b) for $Y_{ff}=\pm m_f/m_W$, for $ f = e, \mu, \tau, b, c, s, d$.}
\label{BRh_a}
\end{figure*}
%

%
\section{Results and discussion}
\label{sec5}

We consider a model with explicit CP violation in the scalar sector, known as 2HDM-III
This model also contains neutral scalar fields that change flavor and have scalar-pseudoscalar interactions with the fermions, as we show in the equation (\ref{yukawa_quarks}). This type of interactions are confronted with the current experimental results, for Higgs decays, through a $\chi^2$ analysis. In this statistical analysis was taken into account the following decay channels : $bb$, $\gamma\gamma$, $ZZ$, $WW$ and $\tau\tau$.  $R_{XX}$ and the branching ratios were obtained in the 2HDM type III with CPV for $h_1$.

From the $\chi^2$ analysis, allowed regions were found for mixing parameters with fixed values of $Y_{ii}$. The results are shown in the Figures \ref{figurey0}, \ref{figurexia1} and \ref{figurexia2}. For large values of $\tan\beta$, it is shown that the allowed region is  significantly suppressed; while in the opposite case, for small or zero values, the regions have a significant increase. This means that large values for $\tan\beta$, which can be consider from 10 approximately, are not viable to observe.

Two extreme cases were also considered for the Yukawa couplings, $Y_{ff}$. One of the cases was to assume $Y_{ff}=0$, while the other case was to consider a maximum value determined by the Cheng-Sher parameterization. Under these assumptions a suppression was obtained when the $Y_{ff}$ couplings participate, as shown in the Figures \ref{figurexia1} and \ref{figurexia2}. The region that shows the greatest suppression is for $\alpha_2$, which is bounded as $|\alpha_2|\lesssim 0.3$.

To study the behavior of the FC parameters, $Y_{ij}$, the reported value for the $b\rightarrow s \gamma$ was considered in an approximate scheme for its branching ratio expression in the 2HDM-III with CPV. Then, allowed values were obtained for $Y_{tc}$ with different values $\tan\beta$ and $Y_{bb}$ by assuming $m_{H^\pm} \approx 500$ GeV. In this scheme the charged Higgs contributes inversely proportional to $\tan\beta$, as it shows the equation (\ref{eq_h}). The values obtained for $Y_{tc}$ are shown in Figure \ref{figurebsg}, which are used to analyze the $t\rightarrow  c Z$ and $t\rightarrow  c h_{1}$ decays in the 2HDM-III with CPV. In the $\Gamma(t\rightarrow  c Z)$, we obtain the analytical expression at the loop level, Eq. (\ref{widthZ}), while for $\Gamma(t\rightarrow  c h_1)$ the expression was calculated at tree level, Eq (\ref{tch}), which are the highest contributions in the model. 

In Figure 7, we show $BR(t\rightarrow  c Z)$  as a function of $\alpha_1$ for different values of $\tan\beta$. The $BR(t\rightarrow  c Z)$ for $Y_{ff} = 0$ can reach values of the order of $10^{-5}$, while in the case $Y_{ff}=m_f/m_W$  the values increase to $8\times10^{-4}$ for $\tan\beta = 1$.

The mixing angle $\alpha_2$ was assigned with random values in the allowed region for SM Higgs and $b\rightarrow  s \gamma$ decays. In figure 7, when we consider $Y_{ff} = 0$ and  the value of $\alpha_1$, in the allowed regions giving by the Higgs decays and $bs\gamma$, the values for $BR (t\rightarrow  c Z)$  are between $10^{-7}$ and $10^{-5}$; but in the limit when $\alpha_1$ goes to zero then  $BR (t\rightarrow  c Z)\sim 10^{-9}$. On the other hand, when $Y_{ff}=m_f/m_W$, the values for $BR (t\rightarrow  c Z)$ are between $10^{-5}$ and $10^{-3}$.

In figure 8, the behavior of $BR (t\rightarrow  c Z)$ as a function of  $\alpha_2$ is analogous to figure 7, but in these cases for $\alpha_2$ allowed regions of the $BR (t\rightarrow  c Z)$ are not suppressed, in fact, in the case of $\alpha_3$ allows the all range and the greatest region for $\alpha_2$. The $\alpha_3$ is set as random values in the interval $\left[ 0,\pi/2\right]$ and the neutral Higgs masses $m_{h_2,h_3}\sim 500$ GeV.

\begin{figure}
\centering 
\includegraphics[scale=0.7]{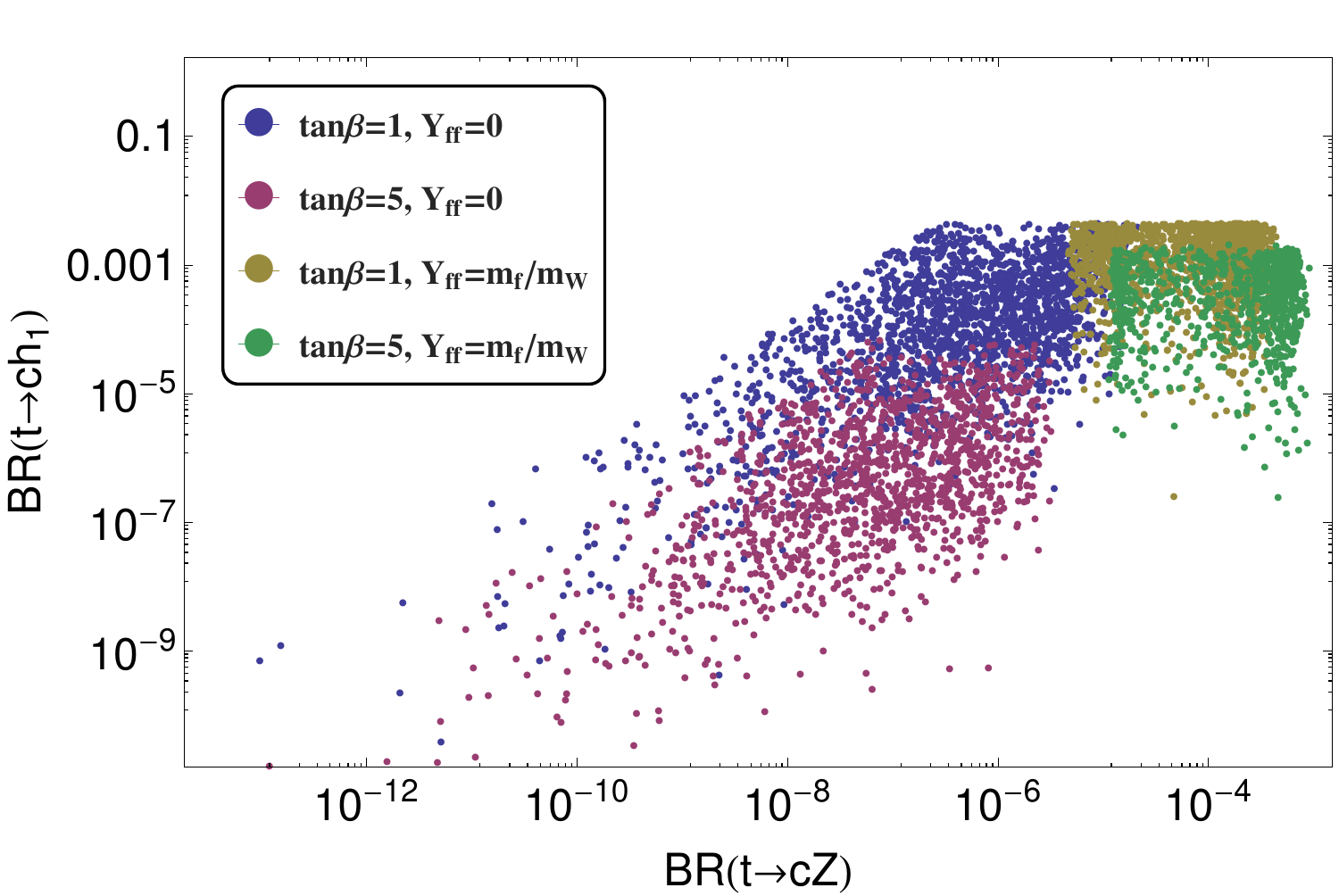}
\caption{\label{BRs} Scatter plot of  $BR(t\rightarrow c Z)$ vs $BR(t\rightarrow c h_1)$ for $\tan\beta=1,5$ and $Y_{ff}=0,\pm m_f/m_W$, for $ f = e, \mu, \tau, b, c, s, d$. We use random values for the allowed region in $\alpha_{1,2}$ found in previous section.}
\end{figure}

For the $BR(t\rightarrow  c h_1)$, the results show a significant increase with respect to the results of $BR (t\rightarrow  c Z)$. For specific regions, as for example $\tan\beta=1$ and $Y_{ff}=m_f/m_W$, the $BR(t\rightarrow  c h_1)$ reaches a maximum value of the order $10^{-2}$. Figure \ref{BRs} shows a scatter plot considering these experimental limits and the allowed regions for the $ \alpha_{1,2}$ parameters that correlate the branching ratios for fixed values of $\tan\beta$ and $Y_{ff}$. This correlation shows a  behavior such that $BR (t\rightarrow  c Z)$ is approximately $10^{2}\sim10^{3}$ times lower than  $BR(t\rightarrow  c h_1)$. This Figure also shows that for $\tan\beta<1$ the values of $BR(t\rightarrow c h_1)$ are close to the reported limit, while for $Y_{ff}=m_f/m_W$ the values of the $BR(t\rightarrow c Z)$ in the model are close to the experimental limit.

The experimental limits are $BR(t\rightarrow c Z)  <  3.7\%$ and $BR(t\rightarrow c H^0) < 0.46\%$~\cite{Patrignani:2016xqp}.  We consider random values for the $\alpha_{1,2}$ parameters in the allowed regions at 90 $\%$ C.L., obtained in the Section 3, to evaluate the $BR(t\rightarrow c Z)$ and $BR(t\rightarrow c h_1)$ in the 2HDM-III with CPV. The evaluation of these branching ratios has also been restricted by experimental limits. Figure \ref{BRs} shows a scatter plot that correlates to  $BR(t\rightarrow  c Z)$ and $BR(t\rightarrow  c h_1)$ for random values of $\alpha_{1,2,3}$ and $Y_{tc}$ in the allowed regions when $\tan\beta$ is set.
%
%
%
\section{Conclusion}
\label{sec6}

We obtain the analytical expressions for $\Gamma(t\rightarrow  c Z)$ and $\Gamma(t\rightarrow  c h_1)$ in the 2HDM type III with CPV at one loop and at tree level. Then, $BR(t\rightarrow  c Z)$ and $BR(t\rightarrow  c h_1)$ are analyzed in allowed regions for $\alpha_1$ and $\alpha_2$ parameters with different fixed values of $\tan\beta$ and Yukawa couplings. These allowed regions are obtained by applying an $\chi^2$ analysis at 90 $\%$ C.L. into the Higgs decays. We also consider the contributions of 2HDM type III to $BR(b\rightarrow s \gamma)$ with its experimental value and we explore values of $Y_{tc}$ as a function of $Y_{bb}$ and $\tan\beta$ for $m_{H^\pm}\approx 500$ GeV. We find feasible scenarios for $\Gamma(t\rightarrow  c Z)$ and $\Gamma(t\rightarrow  c h_1)$  that can be comparable with the current experimental limits.

%
\appendix
\section{Form factors for $t\rightarrow cZ$ decay}
\label{appendix}
The useful notation for masses is introduced as $\hat{x}=\frac{x}{m_t}$. Here, in order to simplify we introduce $\epsilon_{ij}=\frac{1}{2}\left( \sin\beta R_{i1}R_{k3}-\cos\beta R_{i2}R_{k3} \right)$. After the dimensional regularization of the integrals for the $t\rightarrow cZ$ amplitude, we obtain for the Feynman diagrams $D=a,b,c,d$, shown in figure \ref{feynman}, the following results:
\begin{eqnarray}
\label{Va}
\tilde{V}_L^a &=& \int_{0}^{1}dz\int_{0}^{1-z}dy\sum_{i\neq j}\epsilon_{ij}\left(A_{i}+\frac{vY_{tt}}{m_t}B_i\right)B_{j}^{*}\log(\hat{D}_1^2), \nonumber \\
\tilde{V}_R^a &=&\int_{0}^{1}dz\int_{0}^{1-z}dy\sum_{i\neq j}\epsilon_{ij}\left(A_{i}^{*}+\frac{vY_{tt}}{m_t}B_i^{*}\right)B_{j}\log(\hat{D}_1^2).
\end{eqnarray}

\begin{eqnarray}
\label{Fa}
\tilde{F}_L^a &=& \int_{0}^{1}dz\int_{0}^{1-z}dy\sum_{i\neq j}\epsilon_{ij}
\frac{1}{\hat{D}_1^2}\left[(y+z-1)y \left(A_{i}^{*}+\frac{vY_{tt}}{m_t}B_i^{*}\right) B_{j}-y \left(A_{i}+\frac{vY_{tt}}{m_t}B_i\right) B_{j}\right],\nonumber \\
\tilde{F}_R^a &=&\int_{0}^{1}dz\int_{0}^{1-z}dy\sum_{i\neq j}\epsilon_{ij}
\frac{1}{\hat{D}_1^2}\left[(y+z-1)y \left(A_{i}+\frac{vY_{tt}}{m_t}B_i\right) B_{j}^{*}-y \left(A_{i}^{*}+\frac{vY_{tt}}{m_t}B_i^{*}\right) B_{j}^{*}\right].
\end{eqnarray}
\begin{eqnarray}
\label{Vb}
\tilde{V}_L^b &=&\int_{0}^{1}dz\int_{0}^{1-z}dy\sum_{k}\frac{1}{6\hat{D}_2^2}\left[ (3+4s_W^2)\hat{D}_2^2+(3-4s_W^2)(y+z-1)(y+z\hat{M}_Z^2)+8s_W^2\right]\left(A_{k}+\frac{vY_{tt}}{m_t}B_k\right) B_{k}^{*} ,\nonumber \\
\tilde{V}_R^b &=&\int_{0}^{1}dz\int_{0}^{1-z}dy\sum_{k}\frac{1}{6\hat{D}_2^2}\left[ 4s_W^2\hat{D}_2^2+4s_W^2(y+z-1)(y+z\hat{M}_Z^2)+2(3-4s_W^2)\right]\left(A_{k}^{*}+\frac{vY_{tt}}{m_t}B_k^{*}\right)B_{k}.
\end{eqnarray}
\begin{eqnarray}
\label{Fb}
\tilde{F}_L^b &=&\int_{0}^{1}dz\int_{0}^{1-z}dy\sum_{k}\frac{1}{\hat{D}_2^2}\left[(3-4s_W^2)(y-1)A_{k}B_{k}^{*}-4s_W^2(y+z-1)y
\left(A_{k}^{*}+\frac{vY_{tt}}{m_t}B_k^{*}\right)B_{k}\right],\nonumber \\
\tilde{F}_R^b &=&\int_{0}^{1}dz\int_{0}^{1-z}dy\sum_{k}\frac{1}{\hat{D}_2^2}\left[4s_W^2(y-1)A_{k}^{*}B_{k}+(3-4s_W^2)2(y+z-1)y\left(A_{k}+\frac{vY_{tt}}{m_t}B_k\right)B_{k}^{*}\right].
\end{eqnarray}
\begin{eqnarray}
\label{Vc}
\tilde{V}_L^c &=&\int_{0}^{1}dz\int_{0}^{1-z}dy\sum_{k}\frac{1}{6}\log(\hat{D}_3^2)(3-4s_W^2)\left[z\left(A_{k}+\frac{vY_{tt}}{m_t}B_k\right)B_{k}^{*}+\left(A_{k}^{*}+\frac{vY_{tt}}{m_t}B_k^{*}\right)B_{k}\right],\nonumber \\
\tilde{V}_R^c &=&\int_{0}^{1}dz\int_{0}^{1-z}dy\sum_{k}\frac{1}{6}\log(\hat{D}_3^2)(-4s_W^2)\left[z\left(A_{k}^{*}+\frac{vY_{tt}}{m_t}B_k^{*}\right)B_{k}+\left(A_{k}+\frac{vY_{tt}}{m_t}B_k\right)B_{k}^{*}\right].
\end{eqnarray}
\begin{eqnarray}
\label{Vd}
\tilde{V}_L^d &=&\int_{0}^{1}dz\int_{0}^{1-z}dy\sum_{k}\frac{-1}{6}\log(\hat{D}_4^2)(3-4s_W^2)\left(A_{k}^{*}+\frac{vY_{tt}}{m_t}B_k^{*}\right)B_{k}^{*},\nonumber \\
\tilde{V}_R^d &=&\int_{0}^{1}dz\int_{0}^{1-z}dy\sum_{k}\frac{1}{6}\log(\hat{D}_4^2)(4s_W^2)\left(A_{k}+\frac{vY_{tt}}{m_t}B_k\right)B_{k}.
\end{eqnarray}
where
\begin{eqnarray}
\label{ds}
\hat{D}_1^2 &=&(1-\hat{M}_Z^2)yz+\hat{M}_Z^2z^2+(\hat{m}_j-\hat{m}_i-\hat{M}_Z^2)z+y^2-\hat{m}_i^2y+\hat{m}_i^2,\nonumber \\
\hat{D}_2^2 &=&(1-\hat{M}_Z^2)yz+\hat{M}_Z^2(z^2-z)+y^2(\hat{M}_Z^2-2)y+1,\nonumber \\
\hat{D}_3^2 &=&=z^2+(\hat{m}_k^2-2)z+1,\nonumber \\
\hat{D}_4^2 &=&(\hat{m}_k^2-1)z+1.
\end{eqnarray}
\acknowledgments

This work was supported by projects PAPIIT-IN113916 and PAPIIT-IA107118 in DGAPA-UNAM, PIAPIVC07 in FES-Cuautitl\'an UNAM and \emph{Sistema Nacional de Investigadores} (SNI) of the CONACYT in M\'exico. R. Martinez thanks COLCIENCIAS for the financial support. J. H. M. de O. is very grateful for the  comments suggested by Omar G. Miranda to improve the analysis of this work.

%
%

%
%

\begin{thebibliography}{99}
%

\bibitem{Aad:2012tfa} 
  G.~Aad {\it et al.} [ATLAS Collaboration],
  Phys.\ Lett.\ B {\bf 716}, 1 (2012)
  doi:10.1016/j.physletb.2012.08.020
  [arXiv:1207.7214 [hep-ex]].

%
\bibitem{Chatrchyan:2012xdj} 
  S.~Chatrchyan {\it et al.} [CMS Collaboration],
  Phys.\ Lett.\ B {\bf 716}, 30 (2012)
  doi:10.1016/j.physletb.2012.08.021
  [arXiv:1207.7235 [hep-ex]].
  
 
%
\bibitem{Gunion:1989we}
  J.~F.~Gunion, H.~E.~Haber, G.~L.~Kane and S.~Dawson,
  Front.\ Phys.\  {\bf 80}, 1 (2000).


\bibitem{Glashow:1976nt} 
  S.~L.~Glashow and S.~Weinberg,
  Phys.\ Rev.\ D {\bf 15}, 1958 (1977).
  doi:10.1103/PhysRevD.15.1958
 
\bibitem{Atwood:1995ej} 
  D.~Atwood, L.~Reina and A.~Soni,
  Phys.\ Rev.\ Lett.\  {\bf 75}, 3800 (1995)
  doi:10.1103/PhysRevLett.75.3800
  [hep-ph/9507416].

\bibitem{Abbas:2015cua} 
  G.~Abbas, A.~Celis, X.~Q.~Li, J.~Lu and A.~Pich,
  JHEP {\bf 1506}, 005 (2015)
  doi:10.1007/JHEP06(2015)005
  [arXiv:1503.06423 [hep-ph]].

\bibitem{Atwood:1995ud} 
  D.~Atwood, L.~Reina and A.~Soni,
  Phys.\ Rev.\ D {\bf 53}, 1199 (1996)
  doi:10.1103/PhysRevD.53.1199
  [hep-ph/9506243].


\bibitem{Atwood:1996vw} 
  D.~Atwood, L.~Reina and A.~Soni,
  Phys.\ Rev.\ D {\bf 54}, 3296 (1996)
  doi:10.1103/PhysRevD.54.3296
  [hep-ph/9603210].
  
\bibitem{Atwood:1996vj} 
  D.~Atwood, L.~Reina and A.~Soni,
  Phys.\ Rev.\ D {\bf 55}, 3156 (1997)
  doi:10.1103/PhysRevD.55.3156
  [hep-ph/9609279].


\bibitem{Sher:1991km} 
  M.~Sher and Y.~Yuan,
  Phys.\ Rev.\ D {\bf 44}, 1461 (1991).
  doi:10.1103/PhysRevD.44.1461
 
\bibitem{Cheng:1987rs} 
  T.~P.~Cheng and M.~Sher,
  Phys.\ Rev.\ D {\bf 35}, 3484 (1987).
  doi:10.1103/PhysRevD.35.3484


\bibitem{Branco:2011iw}
  G.~C.~Branco, P.~M.~Ferreira, L.~Lavoura, M.~N.~Rebelo, M.~Sher and J.~P.~Silva,
  Phys.\ Rept.\  {\bf 516} (2012) 1
  doi:10.1016/j.physrep.2012.02.002
  [arXiv:1106.0034 [hep-ph]].
 %
\bibitem{Haber:1978jt} 
  H.~E.~Haber, G.~L.~Kane and T.~Sterling,
  Nucl.\ Phys.\ B {\bf 161}, 493 (1979).
  doi:10.1016/0550-3213(79)90225-6
 

\bibitem{Donoghue:1978cj} 
  J.~F.~Donoghue and L.~F.~Li,
  Phys.\ Rev.\ D {\bf 19}, 945 (1979).
  doi:10.1103/PhysRevD.19.945
 
\bibitem{McDonald:1993ex} 
  J.~McDonald,
  Phys.\ Rev.\ D {\bf 50}, 3637 (1994)
  doi:10.1103/PhysRevD.50.3637
  [hep-ph/0702143 [HEP-PH]].
  
\bibitem{McDonald:2001vt} 
  J.~McDonald,
  Phys.\ Rev.\ Lett.\  {\bf 88}, 091304 (2002)
  doi:10.1103/PhysRevLett.88.091304
  [hep-ph/0106249].
  
\bibitem{LopezHonorez:2006gr} 
  L.~Lopez Honorez, E.~Nezri, J.~F.~Oliver and M.~H.~G.~Tytgat,
  JCAP {\bf 0702}, 028 (2007)
  doi:10.1088/1475-7516/2007/02/028
  [hep-ph/0612275].
 
\bibitem{Sakharov:1967dj} 
  A.~D.~Sakharov,
  Pisma Zh.\ Eksp.\ Teor.\ Fiz.\  {\bf 5}, 32 (1967)
  [JETP Lett.\  {\bf 5}, 24 (1967)]
  [Sov.\ Phys.\ Usp.\  {\bf 34}, 392 (1991)]
  [Usp.\ Fiz.\ Nauk {\bf 161}, 61 (1991)].
  doi:10.1070/PU1991v034n05ABEH002497
  
\bibitem{Cosme:2005sb} 
  N.~Cosme, L.~Lopez Honorez and M.~H.~G.~Tytgat,
  Phys.\ Rev.\ D {\bf 72}, 043505 (2005)
  doi:10.1103/PhysRevD.72.043505
  [hep-ph/0506320].
 
\bibitem{Fuyuto:2017ewj}
K. Fuyuto, W. Hou and E. Senaha, [hep-ph/1705.05034].
 
\bibitem{Pontecorvo:1957cp} 
  B.~Pontecorvo,
  Sov.\ Phys.\ JETP {\bf 6}, 429 (1957)
  [Zh.\ Eksp.\ Teor.\ Fiz.\  {\bf 33}, 549 (1957)].
 
\bibitem{Pontecorvo:1957qd} 
  B.~Pontecorvo,
  Sov.\ Phys.\ JETP {\bf 7}, 172 (1958)
  [Zh.\ Eksp.\ Teor.\ Fiz.\  {\bf 34}, 247 (1957)].
 
\bibitem{Maki:1962mu} 
  Z.~Maki, M.~Nakagawa and S.~Sakata,
  Prog.\ Theor.\ Phys.\  {\bf 28}, 870 (1962).
  doi:10.1143/PTP.28.870

\bibitem{Cleveland:1998nv} 
  B.~T.~Cleveland, T.~Daily, R.~Davis, Jr., J.~R.~Distel, K.~Lande, C.~K.~Lee, P.~S.~Wildenhain and J.~Ullman,
  Astrophys.\ J.\  {\bf 496}, 505 (1998).
  doi:10.1086/305343

\bibitem{Fukuda:1996sz} 
  Y.~Fukuda {\it et al.} [Kamiokande Collaboration],
  Phys.\ Rev.\ Lett.\  {\bf 77}, 1683 (1996).
  doi:10.1103/PhysRevLett.77.1683
  
\bibitem{Anselmann:1992um} 
  P.~Anselmann {\it et al.} [GALLEX Collaboration],
  Phys.\ Lett.\ B {\bf 285}, 376 (1992).
  doi:10.1016/0370-2693(92)91521-A

\bibitem{Ahmad:2002jz} 
  Q.~R.~Ahmad {\it et al.} [SNO Collaboration],
  Phys.\ Rev.\ Lett.\  {\bf 89}, 011301 (2002)
  doi:10.1103/PhysRevLett.89.011301
  [nucl-ex/0204008].

\bibitem{Fukuda:2002pe} 
  S.~Fukuda {\it et al.} [Super-Kamiokande Collaboration],
  Phys.\ Lett.\ B {\bf 539}, 179 (2002)
  doi:10.1016/S0370-2693(02)02090-7
  [hep-ex/0205075].

\bibitem{Eguchi:2002dm} 
  K.~Eguchi {\it et al.} [KamLAND Collaboration],
  Phys.\ Rev.\ Lett.\  {\bf 90}, 021802 (2003)
  doi:10.1103/PhysRevLett.90.021802
  [hep-ex/0212021].

\bibitem{Araki:2004mb} 
  T.~Araki {\it et al.} [KamLAND Collaboration],
  Phys.\ Rev.\ Lett.\  {\bf 94}, 081801 (2005)
  doi:10.1103/PhysRevLett.94.081801
  [hep-ex/0406035].

\bibitem{Fukuda:1998mi} 
  Y.~Fukuda {\it et al.} [Super-Kamiokande Collaboration],
  Phys.\ Rev.\ Lett.\  {\bf 81}, 1562 (1998)
  doi:10.1103/PhysRevLett.81.1562
  [hep-ex/9807003].

\bibitem{Ashie:2005ik} 
  Y.~Ashie {\it et al.} [Super-Kamiokande Collaboration],
  Phys.\ Rev.\ D {\bf 71}, 112005 (2005)
  doi:10.1103/PhysRevD.71.112005
  [hep-ex/0501064].

\bibitem{Froggatt:1997he} 
  C.~D.~Froggatt, M.~Gibson and H.~B.~Nielsen,
  Phys.\ Lett.\ B {\bf 409}, 305 (1997)
  doi:10.1016/S0370-2693(97)00934-9
  [hep-ph/9706428].

  

\bibitem{Grossman:1996era} 
  Y.~Grossman,
  Phys.\ Lett.\ B {\bf 380}, 99 (1996)
  doi:10.1016/0370-2693(96)00482-0
  [hep-ph/9603244].
 
\bibitem{Dunietz:2000cr} 
  I.~Dunietz, R.~Fleischer and U.~Nierste,
  Phys.\ Rev.\ D {\bf 63}, 114015 (2001)
  doi:10.1103/PhysRevD.63.114015
  [hep-ph/0012219].
  
  
\bibitem{Langacker:2000ju} 
  P.~Langacker and M.~Plumacher,
  Phys.\ Rev.\ D {\bf 62}, 013006 (2000)
  doi:10.1103/PhysRevD.62.013006
  [hep-ph/0001204].
  
\bibitem{Barger:2003hg} 
  V.~Barger, C.~W.~Chiang, P.~Langacker and H.~S.~Lee,
  Phys.\ Lett.\ B {\bf 580}, 186 (2004)
  doi:10.1016/j.physletb.2003.11.057
  [hep-ph/0310073].
  
  
\bibitem{Fajfer:2001ht} 
  S.~Fajfer and P.~Singer,
  Phys.\ Rev.\ D {\bf 65}, 017301 (2002)
  doi:10.1103/PhysRevD.65.017301
  [hep-ph/0110233].
  
\bibitem{Perez:1992hc} 
  M.~A.~Perez and M.~A.~Soriano,
  Phys.\ Rev.\ D {\bf 46}, 284 (1992).
  doi:10.1103/PhysRevD.46.284
  
\bibitem{Anderson:2005ab} 
  D.~L.~Anderson and M.~Sher,
  Phys.\ Rev.\ D {\bf 72}, 095014 (2005)
  doi:10.1103/PhysRevD.72.095014
  [hep-ph/0509200].
  
\bibitem{Rodriguez:2004mw} 
  J.~A.~Rodriguez and M.~Sher,
  Phys.\ Rev.\ D {\bf 70}, 117702 (2004)
  doi:10.1103/PhysRevD.70.117702
  [hep-ph/0407248].
  
\bibitem{Promberger:2007py} 
  C.~Promberger, S.~Schatt and F.~Schwab,
  Phys.\ Rev.\ D {\bf 75}, 115007 (2007)
  doi:10.1103/PhysRevD.75.115007
  [hep-ph/0702169 [HEP-PH]].
  
\bibitem{Barger:2004qc} 
  V.~Barger, C.~W.~Chiang, J.~Jiang and P.~Langacker,
  Phys.\ Lett.\ B {\bf 596}, 229 (2004)
  doi:10.1016/j.physletb.2004.06.105
  [hep-ph/0405108].
  
\bibitem{Cheung:2006tm} 
  K.~Cheung, C.~W.~Chiang, N.~G.~Deshpande and J.~Jiang,
  Phys.\ Lett.\ B {\bf 652}, 285 (2007)
  doi:10.1016/j.physletb.2007.07.032
  [hep-ph/0604223].
 
 
\bibitem{Grossman:2006ce} 
  Y.~Grossman, Y.~Nir and G.~Raz,
  Phys.\ Rev.\ Lett.\  {\bf 97}, 151801 (2006)
  doi:10.1103/PhysRevLett.97.151801
  [hep-ph/0605028].
 
 
\bibitem{Martinez:2008jj} 
  R.~Martinez and F.~Ochoa,
  Phys.\ Rev.\ D {\bf 77}, 065012 (2008)
  doi:10.1103/PhysRevD.77.065012
  [arXiv:0802.0309 [hep-ph]].
  
\bibitem{Adamson:2016tbq} 
  P.~Adamson {\it et al.} [NOvA Collaboration],
  Phys.\ Rev.\ Lett.\  {\bf 116}, no. 15, 151806 (2016)
  doi:10.1103/PhysRevLett.116.151806
  [arXiv:1601.05022 [hep-ex]].
 
 
\bibitem{GonzalezGarcia:2007ib} 
  M.~C.~Gonzalez-Garcia and M.~Maltoni,
  Phys.\ Rept.\  {\bf 460}, 1 (2008)
  doi:10.1016/j.physrep.2007.12.004
  [arXiv:0704.1800 [hep-ph]].
  
\bibitem{Basso:2012st}
  L.~Basso, A.~Lipniacka, F.~Mahmoudi, S.~Moretti, P.~Osland, G.~M.~Pruna and M.~Purmohammadi,
  JHEP {\bf 1211} (2012) 011
  doi:10.1007/JHEP11(2012)011
  [arXiv:1205.6569 [hep-ph]].
 
\bibitem{Lindner:2016bgg} 
  M.~Lindner, M.~Platscher and F.~S.~Queiroz,
  arXiv:1610.06587 [hep-ph].


\bibitem{Hall:1981bc} 
  L.~J.~Hall and M.~B.~Wise,
  Nucl.\ Phys.\ B {\bf 187}, 397 (1981).
  doi:10.1016/0550-3213(81)90469-7

 
\bibitem{Patrignani:2016xqp} 
  C.~Patrignani {\it et al.} [Particle Data Group],
  Chin.\ Phys.\ C {\bf 40}, no. 10, 100001 (2016).
  doi:10.1088/1674-1137/40/10/100001
 
 
\bibitem{Eilam:1990zc} 
  G.~Eilam, J.~L.~Hewett and A.~Soni,
  Phys.\ Rev.\ D {\bf 44}, 1473 (1991)
  Erratum: [Phys.\ Rev.\ D {\bf 59}, 039901 (1999)].
  doi:10.1103/PhysRevD.44.1473, 10.1103/PhysRevD.59.039901

\bibitem{AguilarSaavedra:2002ns} 
  J.~A.~Aguilar-Saavedra and B.~M.~Nobre,
  Phys.\ Lett.\ B {\bf 553}, 251 (2003)
  doi:10.1016/S0370-2693(02)03230-6
  [hep-ph/0210360].
  
  
\bibitem{AguilarSaavedra:2004wm} 
  J.~A.~Aguilar-Saavedra,
  Acta Phys.\ Polon.\ B {\bf 35}, 2695 (2004)
  [hep-ph/0409342].
  
 
\bibitem{Mele:1998ag} 
  B.~Mele, S.~Petrarca and A.~Soddu,
  Phys.\ Lett.\ B {\bf 435}, 401 (1998)
  doi:10.1016/S0370-2693(98)00822-3
  [hep-ph/9805498].

\bibitem{DiazCruz:1989ub} 
  J.~L.~Diaz-Cruz, R.~Martinez, M.~A.~Perez and A.~Rosado,
  Phys.\ Rev.\ D {\bf 41}, 891 (1990).
  doi:10.1103/PhysRevD.41.891
  
  
\bibitem{Larios:2006pb} 
  F.~Larios, R.~Martinez and M.~A.~Perez,
  Int.\ J.\ Mod.\ Phys.\ A {\bf 21}, 3473 (2006)
  doi:10.1142/S0217751X06033039
  [hep-ph/0605003].
  

%


\bibitem{Grzadkowski:1990sm} 
  B.~Grzadkowski, J.~F.~Gunion and P.~Krawczyk,
  Phys.\ Lett.\ B {\bf 268}, 106 (1991).
  doi:10.1016/0370-2693(91)90931-F
  
\bibitem{Arhrib:2005nx} 
  A.~Arhrib,
  Phys.\ Rev.\ D {\bf 72}, 075016 (2005)
  doi:10.1103/PhysRevD.72.075016
  [hep-ph/0510107].
  
\bibitem{Bejar:2000ub} 
  S.~Bejar, J.~Guasch and J.~Sola,
  Nucl.\ Phys.\ B {\bf 600}, 21 (2001)
  doi:10.1016/S0550-3213(01)00044-X
  [hep-ph/0011091].

\bibitem{Luke:1993cy} 
  M.~E.~Luke and M.~J.~Savage,
  Phys.\ Lett.\ B {\bf 307}, 387 (1993)
  doi:10.1016/0370-2693(93)90238-D
  [hep-ph/9303249].
  

\bibitem{Gaitan:2015hga} 
  R.~Gait\'an, J.~H.~Montes de Oca, E.~A.~Garc\'es and R.~Martinez,
  Phys.\ Rev.\ D {\bf 94}, no. 9, 094038 (2016)
  doi:10.1103/PhysRevD.94.094038
  [arXiv:1503.04391 [hep-ph]].
  
\bibitem{Diaz-Furlong:2016ril} 
  A.~Diaz-Furlong, M.~Frank, N.~Pourtolami, M.~Toharia and R.~Xoxocotzi,
  Phys.\ Rev.\ D {\bf 94}, no. 3, 036001 (2016)
  doi:10.1103/PhysRevD.94.036001
  [arXiv:1603.08929 [hep-ph]].
  
\bibitem{Hesari:2015oya} 
  H.~Hesari, H.~Khanpour and M.~Mohammadi Najafabadi,
  Phys.\ Rev.\ D {\bf 92}, no. 11, 113012 (2015)
  doi:10.1103/PhysRevD.92.113012
  [arXiv:1508.07579 [hep-ph]].
  
  
\bibitem{Enomoto:2015wbn} 
  T.~Enomoto and R.~Watanabe,
  JHEP {\bf 1605}, 002 (2016)
  doi:10.1007/JHEP05(2016)002
  [arXiv:1511.05066 [hep-ph]].
  
\bibitem{Dey:2016cve} 
  U.~K.~Dey and T.~Jha,
  Phys.\ Rev.\ D {\bf 94}, no. 5, 056011 (2016)
  doi:10.1103/PhysRevD.94.056011
  [arXiv:1602.03286 [hep-ph]].
  
  
\bibitem{Khatibi:2015aal} 
  S.~Khatibi and M.~Mohammadi Najafabadi,
  Nucl.\ Phys.\ B {\bf 909}, 607 (2016)
  doi:10.1016/j.nuclphysb.2016.06.009
  [arXiv:1511.00220 [hep-ph]].
  
\bibitem{Gaitan:2017cfa} 
  R.~Gaitan, J.~H.~Montes de Oca and J.~A.~Orduz-Ducuara,
  PTEP {\bf 2017}, no. 7, 073B02 (2017)
  doi:10.1093/ptep/ptx084
  [arXiv:1705.07992 [hep-ph]].
  
\bibitem{Bardhan:2016txk} 
  D.~Bardhan, G.~Bhattacharyya, D.~Ghosh, M.~Patra and S.~Raychaudhuri,
  Phys.\ Rev.\ D {\bf 94}, no. 1, 015026 (2016)
  doi:10.1103/PhysRevD.94.015026
  [arXiv:1601.04165 [hep-ph]].
 
\bibitem{ElKaffas:2006gdt} 
  A.~W.~El Kaffas, W.~Khater, O.~M.~Ogreid and P.~Osland,
  Nucl.\ Phys.\ B {\bf 775}, 45 (2007)
  doi:10.1016/j.nuclphysb.2007.03.041
  [hep-ph/0605142].
 
\bibitem{Ginzburg:2004vp} 
  I.~F.~Ginzburg and M.~Krawczyk,
  Phys.\ Rev.\ D {\bf 72}, 115013 (2005)
  doi:10.1103/PhysRevD.72.115013
  [hep-ph/0408011].

\bibitem{Haber:1993an} 
  H.~E.~Haber and R.~Hempfling,
  Phys.\ Rev.\ D {\bf 48}, 4280 (1993)
  doi:10.1103/PhysRevD.48.4280
  [hep-ph/9307201].

\bibitem{Xu:2017vpq} 
  X.~J.~Xu,
  Phys.\ Rev.\ D {\bf 95}, no. 11, 115019 (2017)
  doi:10.1103/PhysRevD.95.115019
  [arXiv:1705.08965 [hep-ph]].
 
\bibitem{Arhrib:2010ju} 
  A.~Arhrib, E.~Christova, H.~Eberl and E.~Ginina,
  JHEP {\bf 1104}, 089 (2011)
  doi:10.1007/JHEP04(2011)089
  [arXiv:1011.6560 [hep-ph]].
  
\bibitem{Krawczyk:2013jta} 
  M.~Krawczyk, D.~Sokolowska, P.~Swaczyna and B.~Swiezewska,
  JHEP {\bf 1309}, 055 (2013)
  doi:10.1007/JHEP09(2013)055
  [arXiv:1305.6266 [hep-ph]].

\bibitem{Chen:2015gaa} 
  C.~Y.~Chen, S.~Dawson and Y.~Zhang,
  JHEP {\bf 1506}, 056 (2015)
  doi:10.1007/JHEP06(2015)056
  [arXiv:1503.01114 [hep-ph]].
  
\bibitem{Degrassi:2000qf} 
  G.~Degrassi, P.~Gambino and G.~F.~Giudice,
  JHEP {\bf 0012}, 009 (2000)
  doi:10.1088/1126-6708/2000/12/009
  [hep-ph/0009337].
 
\bibitem{Misiak:2006zs} 
  M.~Misiak {\it et al.},
  Phys.\ Rev.\ Lett.\  {\bf 98}, 022002 (2007)
  doi:10.1103/PhysRevLett.98.022002
  [hep-ph/0609232].
 
\bibitem{Lunghi:2006hc} 
  E.~Lunghi and J.~Matias,
  JHEP {\bf 0704}, 058 (2007)
  doi:10.1088/1126-6708/2007/04/058
  [hep-ph/0612166].
  
  
 
\bibitem{Gomez:2006uv} 
  M.~E.~Gomez, T.~Ibrahim, P.~Nath and S.~Skadhauge,
  Phys.\ Rev.\ D {\bf 74}, 015015 (2006)
  doi:10.1103/PhysRevD.74.015015
  [hep-ph/0601163].
  
\bibitem{Barenboim:2013bla} 
  G.~Barenboim, C.~Bosch, M.~L.~Lopez-Iba\~nez and O.~Vives,
  JHEP {\bf 1311}, 051 (2013)
  doi:10.1007/JHEP11(2013)051
  [arXiv:1307.5973 [hep-ph]].

\bibitem{Chen:2001fja} 
  S.~Chen {\it et al.} [CLEO Collaboration],
  Phys.\ Rev.\ Lett.\  {\bf 87}, 251807 (2001)
  doi:10.1103/PhysRevLett.87.251807
  [hep-ex/0108032].
  
\bibitem{Abe:2001hk} 
  K.~Abe {\it et al.} [Belle Collaboration],
  Phys.\ Lett.\ B {\bf 511}, 151 (2001)
  doi:10.1016/S0370-2693(01)00626-8
  [hep-ex/0103042].
  
\bibitem{Lees:2012wg} 
  J.~P.~Lees {\it et al.} [BaBar Collaboration],
  Phys.\ Rev.\ D {\bf 86}, 052012 (2012)
  doi:10.1103/PhysRevD.86.052012
  [arXiv:1207.2520 [hep-ex]].
  
 
\bibitem{Lees:2012ufa} 
  J.~P.~Lees {\it et al.} [BaBar Collaboration],
  Phys.\ Rev.\ D {\bf 86}, 112008 (2012)
  doi:10.1103/PhysRevD.86.112008
  [arXiv:1207.5772 [hep-ex]].
  
\bibitem{Aubert:2007my} 
  B.~Aubert {\it et al.} [BaBar Collaboration],
  Phys.\ Rev.\ D {\bf 77}, 051103 (2008)
  doi:10.1103/PhysRevD.77.051103
  [arXiv:0711.4889 [hep-ex]].
  
  
\bibitem{Amhis:2014hma} 
  Y.~Amhis {\it et al.} [Heavy Flavor Averaging Group (HFAG)],
  arXiv:1412.7515 [hep-ex].
 
 
\bibitem{Kniehl:1994ju} 
  B.~A.~Kniehl and M.~Spira,
  Nucl.\ Phys.\ B {\bf 432}, 39 (1994)
  doi:10.1016/0550-3213(94)90592-4
  [hep-ph/9410319].
 
\bibitem{Djouadi:1995gt} 
  A.~Djouadi, M.~Spira and P.~M.~Zerwas,
  Z.\ Phys.\ C {\bf 70}, 427 (1996)
  doi:10.1007/s002880050120
  [hep-ph/9511344].

\bibitem{Chetyrkin:1995pd} 
  K.~G.~Chetyrkin and A.~Kwiatkowski,
  Nucl.\ Phys.\ B {\bf 461}, 3 (1996)
  doi:10.1016/0550-3213(95)00616-8
  [hep-ph/9505358].
 
 
\bibitem{Larin:1995sq} 
  S.~A.~Larin, T.~van Ritbergen and J.~A.~M.~Vermaseren,
  Phys.\ Lett.\ B {\bf 362}, 134 (1995)
  doi:10.1016/0370-2693(95)01192-S
  [hep-ph/9506465].
 
 
\bibitem{Melnikov:1995yp} 
  K.~Melnikov,
  Phys.\ Rev.\ D {\bf 53}, 5020 (1996)
  doi:10.1103/PhysRevD.53.5020
  [hep-ph/9511310].
  
\bibitem{Chetyrkin:1996sr} 
  K.~G.~Chetyrkin,
  Phys.\ Lett.\ B {\bf 390}, 309 (1997)
  doi:10.1016/S0370-2693(96)01368-8
  [hep-ph/9608318].
  
\bibitem{Spira:2016ztx} 
  M.~Spira,
  Prog.\ Part.\ Nucl.\ Phys.\  {\bf 95}, 98 (2017)
  doi:10.1016/j.ppnp.2017.04.001
  [arXiv:1612.07651 [hep-ph]].
  
\bibitem{Kwiatkowski:1994cu} 
  A.~Kwiatkowski and M.~Steinhauser,
  Phys.\ Lett.\ B {\bf 338}, 66 (1994)
  Erratum: [Phys.\ Lett.\ B {\bf 342}, 455 (1995)]
  doi:10.1016/0370-2693(94)01527-J, 10.1016/0370-2693(94)91345-5
  [hep-ph/9405308].  Erratum: Phys.Lett. B342 (1995) 455
  
  

\bibitem{Djouadi:2005gi} 
  A.~Djouadi,
  Phys.\ Rept.\  {\bf 457}, 1 (2008)
  doi:10.1016/j.physrep.2007.10.004
  [hep-ph/0503172].
  
\bibitem{Spira:1995rr} 
  M.~Spira, A.~Djouadi, D.~Graudenz and P.~M.~Zerwas,
  Nucl.\ Phys.\ B {\bf 453}, 17 (1995)
  doi:10.1016/0550-3213(95)00379-7
  [hep-ph/9504378].
  
\bibitem{Spira:1993bb} 
  M.~Spira, A.~Djouadi, D.~Graudenz and P.~M.~Zerwas,
  Phys.\ Lett.\ B {\bf 318}, 347 (1993).
  doi:10.1016/0370-2693(93)90138-8
 
\bibitem{Djouadi:1990aj} 
  A.~Djouadi, M.~Spira, J.~J.~van der Bij and P.~M.~Zerwas,
  Phys.\ Lett.\ B {\bf 257}, 187 (1991).
  doi:10.1016/0370-2693(91)90879-U
 
\bibitem{Djouadi:1993ji} 
  A.~Djouadi, M.~Spira and P.~M.~Zerwas,
  Phys.\ Lett.\ B {\bf 311}, 255 (1993)
  doi:10.1016/0370-2693(93)90564-X
  [hep-ph/9305335].
 
\bibitem{Liao:1996td} 
  Y.~Liao and X.~y.~Li,
  Phys.\ Lett.\ B {\bf 396}, 225 (1997)
  doi:10.1016/S0370-2693(97)00089-0
  [hep-ph/9605310].
  

\bibitem{Martinez:1989bg} 
  R.~Martinez, M.~A.~Perez and J.~J.~Toscano,
  Phys.\ Rev.\ D {\bf 40}, 1722 (1989).
  doi:10.1103/PhysRevD.40.1722
  
\bibitem{Martinez:1989kr} 
  R.~Martinez, M.~A.~Perez and J.~J.~Toscano,
  Phys.\ Lett.\ B {\bf 234}, 503 (1990).
  doi:10.1016/0370-2693(90)92047-M
 
\bibitem{Martinez:1990ye} 
  R.~Martinez and M.~A.~Perez,
  Nucl.\ Phys.\ B {\bf 347}, 105 (1990).
  doi:10.1016/0550-3213(90)90553-P
 

\bibitem{Spira:1991tj} 
  M.~Spira, A.~Djouadi and P.~M.~Zerwas,
  Phys.\ Lett.\ B {\bf 276}, 350 (1992).
  doi:10.1016/0370-2693(92)90331-W
 
\bibitem{Spira:1997dg} 
  M.~Spira,
  Fortsch.\ Phys.\  {\bf 46}, 203 (1998)
  doi:10.1002/(SICI)1521-3978(199804)46:3<203::AID-PROP203>3.0.CO;2-4
  [hep-ph/9705337].

\bibitem{GonzalezSprinberg:2004bb} 
  G.~A.~Gonzalez-Sprinberg, R.~Martinez and J.~A.~Rodriguez,
  Phys.\ Rev.\ D {\bf 71}, 035003 (2005)
  doi:10.1103/PhysRevD.71.035003
  [hep-ph/0406178].


\bibitem{Cordero-Cid:2013sxa} 
  A.~Cordero-Cid, J.~Hernandez-Sanchez, C.~G.~Honorato, S.~Moretti, M.~A.~Perez and A.~Rosado,
  JHEP {\bf 1407}, 057 (2014)
  doi:10.1007/JHEP07(2014)057
  [arXiv:1312.5614 [hep-ph]].


\bibitem{Keung:1984hn} 
  W.~Y.~Keung and W.~J.~Marciano,
  Phys.\ Rev.\ D {\bf 30}, 248 (1984).
  doi:10.1103/PhysRevD.30.248


\bibitem{deFlorian:2016spz} 
  D.~de Florian {\it et al.} [LHC Higgs Cross Section Working Group],
  doi:10.23731/CYRM-2017-002
  arXiv:1610.07922 [hep-ph].
  
\bibitem{Dabelstein:1991ky} 
  A.~Dabelstein and W.~Hollik,
  Z.\ Phys.\ C {\bf 53}, 507 (1992).
  doi:10.1007/BF01625912
    

\bibitem{Du:1992iy}
  D.~s.~Du and Z.~z.~Xing,
  Phys.\ Rev.\ D {\bf 48} (1993) 2349.
  doi:10.1103/PhysRevD.48.2349

\bibitem{Hall:1993ni} 
  L.~J.~Hall and A.~Rasin,
  Phys.\ Lett.\ B {\bf 315}, 164 (1993)
  doi:10.1016/0370-2693(93)90175-H
  [hep-ph/9303303].
  
\bibitem{Fritzsch:1994yx} 
  H.~Fritzsch and D.~Holtmannspotter,
  Phys.\ Lett.\ B {\bf 338}, 290 (1994)
  doi:10.1016/0370-2693(94)91380-3
  [hep-ph/9406241].

\bibitem{Fritzsch:1995nx} 
  H.~Fritzsch and Z.~z.~Xing,
  Phys.\ Lett.\ B {\bf 353}, 114 (1995)
  doi:10.1016/0370-2693(95)00545-V
  [hep-ph/9502297].

\bibitem{Fritzsch:1997fw} 
  H.~Fritzsch and Z.~Z.~Xing,
  Phys.\ Lett.\ B {\bf 413}, 396 (1997)
  doi:10.1016/S0370-2693(97)01130-1
  [hep-ph/9707215].
  
\bibitem{Fritzsch:1997st} 
  H.~Fritzsch and Z.~z.~Xing,
  Phys.\ Rev.\ D {\bf 57}, 594 (1998)
  doi:10.1103/PhysRevD.57.594
  [hep-ph/9708366].
  
\bibitem{Fritzsch:1999rb} 
  H.~Fritzsch and Z.~z.~Xing,
  Nucl.\ Phys.\ B {\bf 556}, 49 (1999)
  doi:10.1016/S0550-3213(99)00337-5
  [hep-ph/9904286].
  
\bibitem{Branco:1999nb} 
  G.~C.~Branco, D.~Emmanuel-Costa and R.~Gonzalez Felipe,
  Phys.\ Lett.\ B {\bf 477}, 147 (2000)
  doi:10.1016/S0370-2693(00)00193-3
  [hep-ph/9911418].
  
\bibitem{Rosenfeld:2001sc} 
  R.~Rosenfeld and J.~L.~Rosner,
  Phys.\ Lett.\ B {\bf 516}, 408 (2001)
  doi:10.1016/S0370-2693(01)00948-0
  [hep-ph/0106335].
  
\bibitem{Chkareuli:2001dq} 
  J.~L.~Chkareuli, C.~D.~Froggatt and H.~B.~Nielsen,
  Nucl.\ Phys.\ B {\bf 626}, 307 (2002)
  doi:10.1016/S0550-3213(02)00032-9
  [hep-ph/0109156].
  

\bibitem{Fritzsch:2002ga} 
  H.~Fritzsch and Z.~z.~Xing,
  Phys.\ Lett.\ B {\bf 555}, 63 (2003)
  doi:10.1016/S0370-2693(03)00048-0
  [hep-ph/0212195].
 
 
\bibitem{Matsuda:2006xa} 
  K.~Matsuda and H.~Nishiura,
  Phys.\ Rev.\ D {\bf 74}, 033014 (2006)
  doi:10.1103/PhysRevD.74.033014
  [hep-ph/0606142].
  
\bibitem{Koide:2002cj} 
  Y.~Koide, H.~Nishiura, K.~Matsuda, T.~Kikuchi and T.~Fukuyama,
  Phys.\ Rev.\ D {\bf 66}, 093006 (2002)
  doi:10.1103/PhysRevD.66.093006
  [hep-ph/0209333].
  
\bibitem{Matsuda:2003zm} 
  K.~Matsuda and H.~Nishiura,
  Phys.\ Rev.\ D {\bf 69}, 053005 (2004)
  doi:10.1103/PhysRevD.69.053005
  [hep-ph/0309272].
  
  
\bibitem{CarcamoHernandez:2005ka} 
  A.~E.~Carcamo Hernandez, R.~Martinez and F.~Ochoa,
  Phys.\ Rev.\ D {\bf 73}, 035007 (2006)
  doi:10.1103/PhysRevD.73.035007
  [hep-ph/0510421].
  
\bibitem{Crivellin:2013wna} 
  A.~Crivellin, A.~Kokulu and C.~Greub,
  Phys.\ Rev.\ D {\bf 87}, no. 9, 094031 (2013)
  doi:10.1103/PhysRevD.87.094031
  [arXiv:1303.5877 [hep-ph]].
  
\bibitem{Brod:2013cka} 
  J.~Brod, U.~Haisch and J.~Zupan,
  JHEP {\bf 1311}, 180 (2013)
  doi:10.1007/JHEP11(2013)180
  [arXiv:1310.1385 [hep-ph]].

  
\bibitem{Gorbahn:2014sha} 
  M.~Gorbahn and U.~Haisch,
  JHEP {\bf 1406}, 033 (2014)
  doi:10.1007/JHEP06(2014)033
  [arXiv:1404.4873 [hep-ph]].

\end{thebibliography}
\end{document}